\documentclass[12pt]{article}
\usepackage{amssymb,amsmath,epsfig}
\begin{document} \parindent=0pt
\parskip=6pt \rm

\begin{center}

{\bf \Large  Theory of ferromagnetic unconventional superconductors with spin-triplet electron pairing}

\vspace{0.3cm}

{\bf  Dimo I. Uzunov}

Collective Phenomena Laboratory, G. Nadjakov Institute of Solid State Physics,\\ Bulgarian
Academy of Sciences, BG-1784 Sofia, Bulgaria.\\

\end{center}

{\bf Pacs}: 74.20.De, 74.25.Dw, 64.70.Tg\\

electrons, multi-critical point,
phase diagram.

\begin{abstract}

A general phenomenological theory is presented for the phase behavior of ferromagnetic superconductors with
spin-triplet electron Cooper pairing. The theory describes in details the temperature-pressure phase
diagrams of real intermetallic compounds exhibiting the remarkable phenomenon of coexistence of
spontaneous magnetic moment of the itinerant electrons and spin-triplet superconductivity.
The quantum phase transitions which may occur in these systems are also
described. The theory allows for a classification of these itinerant ferromagnetic superconductors in
two types: type I and type II. The classification is based on quantitative criteria.The comparison of
theory and experiment is performed and outstanding problems are discussed.
\end{abstract}

\section{Introduction}

In the beginning of this century the unconventional superconductivity of spin-triplet type had
been experimentally discovered in several itinerant ferromagnets. Since then much
experimental and theoretical research on the properties of these systems has been accomplished.
Here we review the phenomenological theory of ferromagnetic
unconventional superconductors with spin-triplet Cooper pairing of electrons.
Some theoretical aspects of the description of the phases and the
phase transitions in these interesting systems, including the remarkable phenomenon
of coexistence of superconductivity and ferromagnetism are discussed with an
emphasis on the comparison of theoretical results with experimental data.

The spin-triplet  or $p$-wave pairing allows parallel spin
orientation of the fermion Cooper pairs in superfluid $^3$He and
unconventional superconductors~\cite{Vollhardt:1990}. For this
reason the resulting unconventional superconductivity is robust
with respect to effects of external magnetic field and spontaneous
ferromagnetic ordering, so it may coexist with the latter. This
general argument implies that there could be metallic compounds
and alloys, for which the  coexistence of spin-triplet
superconductivity and ferromagnetism may be observed.

Particularly, both superconductivity and itinerant ferromagnetic
orders can be created by the same band electrons in the metal,
which means that spin-1 electron Cooper pairs participate in the formation
of the itinerant ferromagnetic order. Moreover, under certain
conditions the superconductivity is enhanced rather than
depressed by the uniform ferromagnetic order that can generate
it, even in cases when the superconductivity does not appear in a pure form
as a net result of indirect electron-electron coupling.

The coexistence of superconductivity and ferromagnetism as a result of collective
behavior of  $f$-band electrons has been found experimentally for
some Uranium-based intermetallic compounds as,
UGe$_2$~\cite{Saxena:2000,Huxley:2001, Tateiwa:2001, Harada:2007},
URhGe~\cite{Aoki:2001, Hardy1:2005, Hardy2:2005},
UCoGe~\cite{Huy:2007, Huy:2008}, and UIr~\cite{Akazawa:2005,
Kobayashi:2006}. At low temperature $(T \sim 1~\mbox{K})$ all
these compounds exhibit thermodynamically stable phase of
coexistence of spin-triplet superconductivity and itinerant
($f$-band) electron ferromagnetism (in short, FS phase). In
UGe$_2$ and UIr the FS phase appears at high pressure $(P \sim
1~\mbox{GPa})$ whereas in URhGe and UCoGe,  the coexistence  phase
persists up to ambient pressure $(10^5 \mbox{Pa} \equiv 1
\mbox{bar})$.

Experiments, carried out in ZrZn$_2$ ~\cite{Pfleiderer:2001},
also indicated the appearance of  FS phase at $ T < 1$ K in a
wide range of pressures ($0<P \sim 21~\mbox{kbar}$). In Zr-based
compounds the ferromagnetism and the $p$-wave superconductivity
occur as a result of the collective behavior of the $d$-band
electrons. Later experimental results~\cite{Yelland1:2005,
Yelland2:2005} had imposed the conclusion that bulk
superconductivity is lacking in ZrZn$_2$, but the occurrence of a
surface FS phase at surfaces with higher Zr content than that in
ZrZn$_2$ has been reliably demonstrated. Thus the problem for the
coexistence of bulk superconductivity with ferromagnetism in
ZrZn$_2$ is still unresolved. This raises the question whether the
FS phase in ZrZn$_2$ should be studied by surface thermodynamics
methods or should it be investigated by considering that bulk and
surface thermodynamic phenomena can be treated on the same
footing. Taking into account the mentioned experimental results for
ZrZn$_2$ and their interpretation by the
experimentalists~\cite{Pfleiderer:2001,Yelland1:2005,
Yelland2:2005} we assume that the unified thermodynamic approach
can be applied. As an argument supporting this point of view let us mention that
the spin-triplet superconductivity occurs not only in bulk materials but also in
quasi-two-dimensional (2D) systems -- thin films and surfaces and quasi-1D wires
(see, e.g., Refs.~\cite{Bolesh:2005}). In ZrZn$_2$ and UGe$_2$ both
ferromagnetic and superconducting orders vanish at the
same critical pressure $P_c$, a fact implying that the respective order
parameter fields strongly depend on each other and should be
studied on the same thermodynamic basis~\cite{Nevidomskyy:2005}.

\begin{figure}
\includegraphics[width=8cm, height=6cm, angle=0]{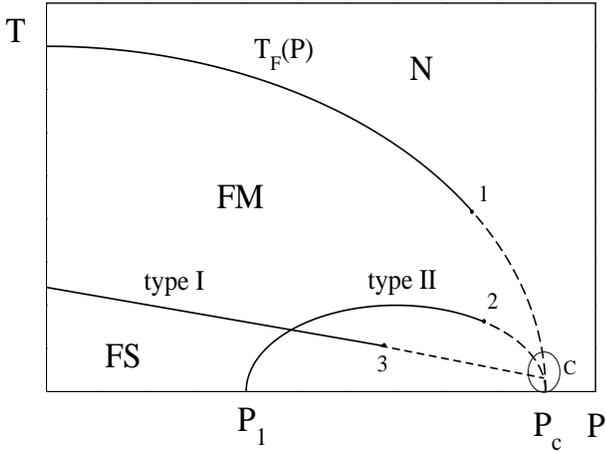}
\caption{\label{Fig1} \small  An illustration of $T-P$ phase diagram
of $p$-wave ferromagnetic superconductors (details are omitted): N -- normal phase,
FM -- ferromagnetic phase, FS -- phase of coexistence of ferromagnetic
order and superconductivity, $T_{F}(P)$ and $T_{FS}(P)$ are the
respective phase transition lines: solid lines correspond to
second order phase transitions, dashed lines stand for first
order phase transition; $1$ and $2$ are tricritical points;
$P_c$ is the critical pressure, and the circle $C$ surrounds
a relatively small domain of high pressure and
low temperature, where the phase diagram may have several forms depending on
the particular substance. The line of the FM-FS phase transition may
extend up to ambient pressure (type I ferromagnetic superconductors),
or, may terminate at $T=0$ at some high pressure $P=P_1$
(type II ferromagnetic superconductors, as indicated in the figure). }
\end{figure}

Fig.~\ref{Fig1} illustrates the shape of the $T-P$ phase diagrams of real
intermetallic compounds. The phase transition from the normal (N)
to the ferromagnetic phase (FM) (in short, N-FM transition)
is shown by the line $T_F(P)$. The line $T_{FS}(P)$ of the phase transition from FM
to FS (FM-FS transition) may have two or more distinct shapes. Beginning
from the maximal (critical) pressure $P_c$, this line may extend, like in ZrZn$_2$,
to all pressures $P < P_c$, including the ambient pressure $P_a$;
see the almost straight line containing the point 3 in
Fig.~\ref{Fig1}. A second possible form of this line, as known, for example,
from UGe$_2$ experiments, is shown in Fig.~\ref{Fig1}
by the curve which begins at $P \sim P_c$, passes through the
point 2, and terminates at some pressure $P_{1} > P_a$, where the
superconductivity vanishes. These are two qualitatively different
physical pictures: (a) when the superconductivity survives up to
ambient pressure (type I), and (b) when the superconducting states are
possible only at relatively high pressure (for UGe$_2$, $P_1 \sim
1$ GPa); type II. At the tricritical points 1, 2 and
3 the order of the phase transitions changes from second order (solid lines) to first order
(dashed lines). It should be emphasized that in all compounds, mentioned above,  $T_{FS}(P)$ is much lower than
$T_{F}(P)$ when the pressure $P$ is considerably below the critical pressure $P_c$ (for experimental data, see Sec.~8).

In Fig.~\ref{Fig1}, the circle $C$ denotes a narrow domain around
$P_c$ at relatively low temperatures ($T \lesssim 300$ mK), where
the experimental data are quite few and the predictions about the shape of
the phase transition are not reliable. It could be assumed, as in the
most part of the experimental papers, that $(T=0,P=P_c)$ is the zero
temperature point at which both lines $T_{F}(P)$ and $T_{FS}(P)$ terminate.
A second possibility is that these lines may join in a single (N-FS)
phase transition line at some point $(T\gtrsim 0,
P^{\prime}_c\lesssim P_c)$ above the absolute zero. In this second
variant, a direct N-FS phase transition occurs, although this
option exists in a very small domain of temperature and pressure variations:
from point $(0,P_c)$ to point  $(T\gtrsim 0,
P^{\prime}_c\lesssim P_c)$. A third variant is related with the
possible splitting of the point $(0,P_c)$, so that the N-FM line
terminates at $(0,P_c)$, whereas the FM-FS line terminates at
another zero temperature point $(0, P_{0c})$; $P_{0c} \lesssim P_c
$. In this case, the $p$-wave ferromagnetic superconductor has
three points of quantum (zero temperature) phase
transitions~\cite{Cottam:2008, Shopova:2009}.

These and other possible shapes of  $T-P$ phase diagrams are described
within the framework of the general theory of Ginzburg-Landau (GL)
type~\cite{Cottam:2008, Shopova:2009,Shopova:2005} in a conformity with
the experimental data; see also Ref.~\cite{Shopova:2006}. The same theory has been
confirmed by a microscopic derivation based on a microscopic Hamiltonian including a
spin-generalized BCS term and an additional Heisenberg exchange term~\cite{Dahl:2007}.

For all compounds, cited above, the FS phase occurs only in the
ferromagnetic phase domain of the $T-P$ diagram. Particularly at
equilibrium, and for  given $P$, the temperature $T_{F}(P)$ of the
normal-to-ferromagnetic phase (or N-FM) transition is never lower
than the temperature $T_{FS}(P)$ of the ferromagnetic-to-FS phase
transition (FM-FS transition). This confirms the point of view
that the superconductivity in these compounds is triggered by the
spontaneous magnetization $\mbox{\boldmath$M$}$, in analogy with
the well-known triggering of the superfluid phase A$_1$ in $^3$He
at mK temperatures by the external magnetic field
$\mbox{\boldmath$H$}$. Such ``helium analogy" has been used in
some theoretical studies (see, e.g., Ref.~\cite{Machida:2001,
Walker:2002}), where Ginzburg-Landau (GL) free energy terms,
describing the FS phase were derived by symmetry group arguments.
The non-unitary state, with a non-zero value of the Cooper pair
magnetic moment, known from the theory of unconventional
superconductors and superfluidity in $^3$He~\cite{Vollhardt:1990},
has been suggested firstly in Ref.~\cite{Machida:2001}, and later
confirmed in other studies~\cite{Hardy1:2005, Walker:2002};
recently, the same topic was comprehensively discussed
in Ref.~\cite{Linder:2007}.

For the spin-triplet ferromagnetic superconductors the trigger
mechanism was recently examined in detail~\cite{Shopova:2005,
Shopova:2006}. The system main properties are specified by
terms in the GL expansion of form $M_i\psi_j\psi_k$, which represent the
interaction of the magnetization $\mbox{\boldmath$M$} = \{M_j; j=1,2,3\}$ with the complex
superconducting vector field $\mbox{\boldmath$\psi$} =\{\psi_j; j=1,2,3\}$.
Particularly, these terms are responsible for the appearance of superconductivity
($|\mbox{\boldmath$\psi$}| > 0$) for certain $T$ and $P$ values. A
similar trigger mechanism is familiar in the context of improper
ferroelectrics~\cite{Cowley:1980}.

A crucial feature of these systems is the nonzero magnetic moment
of the spin-triplet Cooper pairs. As mentioned above, the
microscopic theory of magnetism and superconductivity in non-Fermi
liquids of strongly interacting heavy electrons ($f$ and $d$ band
electrons) is either too complex or insufficiently developed to
describe the complicated behavior in itinerant ferromagnetic
compounds. Several authors (see~\cite{Shopova:2005, Shopova:2006, Machida:2001,
Walker:2002, Linder:2007}) have explored the phenomenological
description by a self-consistent mean field theory, and here we will
essentially use the thermodynamic results, in particular, results from the
analysis in Refs.~\cite{Shopova:2005, Shopova:2006}. Mean-field
microscopic theory of spin-mediated pairing leading to the
mentioned non-unitary superconductivity state has been developed
in Ref.~\cite{Nevidomskyy:2005} that is in conformity with the
phenomenological description that we have done.

The coexistence of $s$-wave (conventional) superconductivity and ferromagnetic order is a long-standing problem in
condensed matter physics~\cite{Vonsovskii:1982, Bulaevskii:1984, Blount:1979}. While the $s$-state
Cooper pairs contain only opposite electron spins and  can easily
be destroyed by the spontaneous magnetic moment, the spin-triplet Cooper pairs possess quantum states
with parallel orientation of the electron spins and therefore can survive in the presence of
substantial magnetic moments. This is the basic difference in the magnetic behavior of
conventional ($s$-state) and spin-triplet superconductivity phases.
In contrast to other superconducting materials, for example, ternaty and Chevrel phase compounds,
where the effect of magnetic order on $s$-wave superconductivity has
been intensively studied in the seventies and eighties of last century (see, e.g.,
Refs.~\cite{Vonsovskii:1982, Bulaevskii:1984, Blount:1979}),
in these ferromagnetic compounds the phase transition
temperature $T_{F}$ to the ferromagnetic state is much higher than the phase transition
temperature $T_{FS}$ from ferromagnetic to a (mixed) state of coexistence of ferromagnetism
and superconductivity. For example, in UGe$_2$ we have $T_{FS} \sim 0.8$ K versus
maximal $T_{F} = 52$ K~\cite{Saxena:2000,Huxley:2001, Tateiwa:2001, Harada:2007}.
Another important difference between the ternary rare earth compounds and the intermetallic compounds
(UGe$_2$, UCoGe, etc.), which are of interest in this paper, is that the experiments with the latter do no give any evidence
for the existence of a standard normal-to-superconducting phase
transition in zero external magnetic field. This is an indication that the (generic) critical temperature $T_s$
of the pure superconductivity state in these intermetallic compounds is very low ($T_s \ll T_{FS}$), if not zero or even negative.

In the reminder of this paper, we present general thermodynamic treatment of
systems with itinerant ferromagnetic order and superconductivity
due to spin-triplet Cooper pairing of the same band electrons,
which are responsible for the spontaneous magnetic moment.
The usual Ginzburg-Landau (GL) theory of superconductors has been completed
to include the complexity of the vector order parameter $\mbox{\boldmath$\psi$}$,
the magnetization $\mbox{\boldmath$M$}$ and new relevant energy
terms~\cite{Shopova:2005, Shopova:2006}. We outline the $T-P$
phase diagrams of ferromagnetic spin-triplet
superconductors and demonstrate that in these materials two contrasting
types of thermodynamic behavior are possible. The present phenomenological
approach includes both mean-field and spin-fluctuation theory
(SFT), as the arguments in Ref.~\cite{Yamada:1993}. We propose a
simple, yet comprehensive, modeling of  $P$ dependence of the
free energy parameters, resulting in a very good compliance of
our theoretical predictions for the shape the $T-P$ phase diagrams
with the experimental data (for some preliminary results, see
Ref.~\cite{Cottam:2008, Shopova:2009}).

The theoretical analysis is done by the standard methods of phase transition
theory~\cite{Uzunov:1993}. Treatment of fluctuation effects and quantum
correlations~\cite{Uzunov:1993, Shopova:2003} is not included in this study.
But the parameters of the generalized GL free energy may be considered
either in mean-field approximation as here, or  as phenomenologically
renormalized parameters which are affected by additional physical
phenomena, as for example, spin fluctuations.

We demonstrate with the help of present theory that we can
outline different possible topologies for the $T-P$ phase diagram,
depending on the values of Landau parameters, derived from the
existing  experimental data. We show that for spin-triplet ferromagnetic
superconductors there exist two distinct types of behavior, which we denote as
Zr-type (or, alternatively, type I) and U-type (or, type II); see Fig.~\ref{Fig1}.
This classification of the FS, first mentioned in Ref.~\cite{Cottam:2008}, is based
on the reliable interrelationship between a quantitative criterion derived by us
and the thermodynamic properties of the ferromagnetic spin-triplet superconductors.
Our approach can be also applied to URhGe, UCoGe, and UIr. The results shed light on
the problems connected with the order of the quantum phase transitions at ultra-low and
zero temperatures. They also raise the question for further experimental investigations
of the detailed structure of the phase diagrams in the high-$P$/low-$T$ region.

\section{Theoretical framework}

Consider the GL free energy functional of the form
\begin{equation}
\label{Eq1}
F(\mbox{\boldmath$\psi$}, \mbox{\boldmath$B$}) = \int_V d \mbox{\boldmath$x$}
\left[ f_{\mbox{\scriptsize S}}(\mbox{\boldmath$\psi$}) +
f_{\mbox{\scriptsize
F}}(\mbox{\boldmath$M$}) +
f_{\mbox{\scriptsize I}}(\psi,\mbox{\boldmath$M$}) +
\frac{\mbox{\boldmath$B$}^2}{8\pi} - \mbox{\boldmath$B.M$} \right],
\end{equation}
where the fields $\mbox{\boldmath$\psi$}$, $\mbox{\boldmath$M$}$,
and $\mbox{\boldmath$B$}$ are supposed to depend on the spatial vector
$\mbox{\boldmath$x$} \in V$ in the volume $V$ of the superconductor.
In Eq.~(\ref{Eq1}), the free energy density generated by the generic
superconducting subsystem ($\mbox{\boldmath$\psi$}$) is given by
\begin{equation}
\label{Eq2}
f_{\mbox{\scriptsize S}}(\mbox{\boldmath$\psi$})= f_{grad}(\mbox{\boldmath$\psi$})
 + a_s|\mbox{\boldmath$\psi$}|^2 +\frac{b_s}{2}|\mbox{\boldmath$\psi$}|^4 +
 \frac{u_s}{2}|\mbox{\boldmath$\psi$}^2|^2 +
\frac{v_s}{2}\sum_{j=1}^{3}|\psi_j|^4 \;,
\end{equation}
with
\begin{eqnarray}
\label{Eq3}
f_{grad}(\mbox{\boldmath$\psi$})&  =
& K_1(D_i\psi_j)^{\ast}(D_iD_j) +K_2\left[
 (D_i\psi_i)^{\ast}(D_j\psi_j) + (D_i\psi_j)^{\ast}(D_j\psi_i)\right] \\
\nonumber
&& + K_3(D_i\psi_i)^{\ast}(D_i\psi_i),
\end{eqnarray}
where a summation over the indices $(i,j)$  is assumed, the symbol $D_j
= (\hbar\partial/i\partial x_j + 2|e|A_j/c)$
of covariant differentiation is introduced, and $K_j$ are material parameters~\cite{Vollhardt:1990}.
The free energy density $f_{\mbox{\scriptsize F}}(\mbox{\boldmath$M$})$  of
a standard ferromagnetic phase transition of second order~\cite{Uzunov:1993}  is
\begin{equation}
\label{Eq4}
f_{\mbox{\scriptsize F}}(\mbox{\boldmath$M$}) =
c_f\sum_{j=1}^{3}|\nabla\mbox{\boldmath$M$}_j|^2 +
 a_f\mbox{\boldmath$M$}^2 + \frac{b_f}{2}\mbox{\boldmath$M$}^4,
\end{equation}
with $c_f, b_f > 0$, and $a_f = \alpha (T-T_f)$, where $\alpha_f>0$ and $T_f$
is the critical temperature, corresponding of the generic ferromagnetic phase transition.
Finally, the energy $f_{\mbox{\scriptsize I}}(\mbox{\boldmath$\psi$}, \mbox{\boldmath$M$})$
produced by the possible couplings of $\mbox{\boldmath$\psi$}$ and $\mbox{\boldmath$M$}$
is given by
\begin{equation}
\label{Eq5}
f_{\mbox{\scriptsize I}}(\mbox{\boldmath$\psi$}, \mbox{\boldmath$M$}) = i\gamma_0
\mbox{\boldmath$M$}.(\mbox{\boldmath$\psi$}\times \mbox{\boldmath$\psi$}^*) +
\delta_0 \mbox{\boldmath$M$}^2|\mbox{\boldmath$\psi$}|^2,
\end{equation}
where the coupling parameter $\gamma_0 \sim J$ depends on the ferromagnetic
exchange parameter $J>0$,~\cite{Machida:2001, Walker:2002} and $\delta_0$ is the standard
$\mbox{\boldmath$M$}-\mbox{\boldmath$\psi$}$ coupling parameter, known from the theory
of multicritical phenomena~\cite{Uzunov:1993} and from studies of coexistence of ferromagnetism and
superconductivity in ternary compounds~\cite{Vonsovskii:1982, Bulaevskii:1984}.

As usual, in Eq.~(\ref{Eq2}), $a_s = (T-T_s)$, where $T_s$ is the critical
temperature of the generic superconducting transition,
$b_s > 0$. The parameters $u_s$ and $v_s$ and $\delta_0$ may take
some negative values, provided the overall stability of
the system is preserved. The values of the material parameters
$\mu$ = ($T_s$, $T_f$, $\alpha_s$, $\alpha_f$, $b_s$, $u_s$, $v_s$, $b_f$, $K_j$,
$\gamma_0$ and $\delta_0$) depend on the choice of the substance and on intensive
thermodynamic parameters, such as the temperature $T$ and the pressure $P$.
From a microscopic point of view, the parameters $\mu$  depend on the density of
states $U_F(k_F)$ on the Fermi surface. On the other hand $U_F$ varies with $T$ and $P$.
Thus the relationships $(T,P) \rightleftarrows U_F \rightleftarrows \mu$, i.e.,
the functional relations $\mu[U_F(T,P)]$, are of essential interest.
While these relations are unknown, one may suppose some direct dependence $\mu(T,P)$.
The latter should correspond to the experimental data.

The free energy (\ref{Eq1}) is quite general. It has been deduced by several reliable
arguments. In order to construct Eq.~(\ref{Eq1})--(\ref{Eq5}) we have used the
standard GL theory of superconductors and the phase transition theory with an account
of the relevant anisotropy of the $p$-wave Cooper pairs and the crystal anisotropy,
described by the $u_s$- and $v_s$-terms in Eq.~(\ref{Eq2}), respectively. Besides,
we have used the general case of cubic anisotropy, when all three components
$\psi_j$ of $\mbox{\boldmath$\psi$}$ are relevant. Note, that in certain real cases,
for example, in UGe$_2$, the crystal symmetry is tetragonal, $\mbox{\boldmath$\psi$}$  effectively
behaves as a two-component vector and this leads to a considerable simplification of the
theory. As shown in Ref.~\cite{Shopova:2005}, the mentioned anisotropy terms are not
essential in the description of the main thermodynamic properties, including the shape
of the $T-P$ phase diagram. For this reason we shall often ignore the respective
terms in Eq.~(\ref{Eq2}). The $\gamma_0$-term triggers the superconductivity
($\mbox{\boldmath$M$}$-triger effect~\cite{Shopova:2005, Shopova:2006}) while the
$\delta_0\mbox{\boldmath$M$}^2 |\psi|^2$--term makes the model more
realistic for large values of $\mbox{\boldmath$M$}$. This allows for an extension of
the domain of the stable ferromagnetic order up to zero temperatures for a wide range
of values of the material parameters and the pressure $P$. Such a picture corresponds
to the real situation in ferromagnetic compounds~\cite{Shopova:2005}.

The total free energy (\ref{Eq1}) is difficult for a theoretical
investigation. The various vortex and uniform phases described by
this complex model cannot be investigated within a single
calculation but rather one should focus on particular problems. In
Ref.~\cite{Walker:2002} the vortex phase was discussed with the
help of the criterion~\cite{Abrikosov:1957} for a stability of
this state near the phase transition line $T_{c2}(\mbox{\boldmath$B$})$, ; see also,
Ref.~\cite{Lifshitz:1980}. The phase transition line $T_{c2}(H)$ of a usual superconductor
in external magnetic field $H = |\mbox{\boldmath$H$}|$ is located above the phase transition line $T_s$ of
the uniform (Meissner) phase. The reason is that $T_s$ is defined by the equation $a_s(T) = 0$,
whereas $T_{c2}(H)$ is a solution of the equation $|a_s| = \mu_BH$, where
$\mu_B = |e|\hbar/2mc$ is the Bohr magneton~\cite{Lifshitz:1980}. For ferromagnetic superconductors,
where $M > 0$, one should use the magnetic induction $\mbox{\boldmath$B$}$ rather than \mbox{\boldmath$H$}.
In case of $\mbox{\boldmath$H$} = 0$ one should apply the
same criterion with respect to the magnetization $\mbox{\boldmath$M$}$ for
small values of $|\mbox{\boldmath$\psi$}|$ near the phase transition line
$T_{c2}(M)$; $M = |\mbox{\boldmath$M$}|$. For this reason we should use the
diagonal quadratic form~\cite{Belich:2010} corresponding to the entire
$\mbox{\boldmath$\psi$}^2$-part of the total free energy functional (\ref{Eq1}).
The lowest energy term in this diagonal quadratic part contains a coefficient $a$ of the
form $a = (a_s - \gamma_0 M - \delta M^2)$~\cite{Belich:2010}. Now the equation $a(T) = 0$
defines the critical temperature of the Meissner phase and the equation $|a_s| = \mu_BM$ stands for $T_{c2}(M)$.
It is readily seen that these two equations can be written in the same form, provided the parameter
$\gamma_0$ in $a$ is substituted by   $\gamma_0^{\prime} = (\gamma_0- \mu_B$). Thus the
phase transition line corresponding to the vortex phase, described by the model (\ref{Eq1}) at
zero external magnetic field and generated by the magnetization $\mbox{\boldmath$M$}$, can be obtained
from the phase transition line corresponding to the uniform superconducting phase by an effective change
of the value of the parameter $\gamma_0$. Both lines have the same shape and this is a particular
property of the present model. The variation of the parameter $\gamma_0$ generates a family of lines.

Now we propose a possible way of theoretical treatment of the $T_{FS}(P)$ line of the FM-FS
phase transition, shown in Fig.~(1). This is a crucial point in our theory. The phase transition line of the uniform superconducting phase can
be calculated within the thermodynamic analysis of the uniform phases, described by
the free energy (\ref{Eq1}). This analysis is done in a simple variant of
the free energy (\ref{Eq1}) in which the fields $\mbox{\boldmath$\psi$}$ and $\mbox{\boldmath$M$}$
do not depend on the spatial vector $\mbox{\boldmath$x$}$. The accomplishment of such analysis will
give a formula for the phase transition line $T_{FS}(P)$ which corresponds a Meissner phase coexisting
with the ferromagnetic order. The theoretical result for $T_{FS}(P)$ will contain
a unspecified parameter $\gamma_0$. If the theoretical line $T_{FS}(P)$  is fitted to the experimental
data for the FM-FS transition line corresponding to a particular compound, the two curves will coincide for some value of
$\gamma_0$, irrespectively on the structure of the FS phase. If the FS phase contains a vortex superconductivity
the fitting parameter $\gamma_{0(eff)}$ should be interpreted as $\gamma_0^{\prime}$
but if the FS phase contains Meissner superconductivity, $\gamma_{0(eff)}$ should be identified as $\gamma_0$.
These arguments justify our approach to the investigation of the experimental data for the phase diagrams of
intermetallic compounds with FM and FS phases. In the remainder of this paper, we shall investigate uniform phases.

\section{Model considerations}

In the previous section we have justified a thermodynamic analysis of the free energy (\ref{Eq1}) in terms of uniform order parameters.
Neglecting the $\mbox{\boldmath$x$}$-dependence of $\mbox{\boldmath$\psi$}$ and
$\mbox{\boldmath$M$}$, the free energy per unit volume,
$F/V =f(\mbox{\boldmath$\psi$},\mbox{\boldmath$M$})$ in zero external magnetic field
$(\mbox{\boldmath$H$}=0)$, can be written in the form

\begin{eqnarray}
\label{Eq6} f(\mbox{\boldmath$\psi$},\mbox{\boldmath$M$}) &= &
  a_s|\mbox{\boldmath$\psi$}|^2 +\frac{b_s}{2}|\mbox{\boldmath$\psi$}|^4 +
  \frac{u_s}{2}|\mbox{\boldmath$\psi$}^2|^2 +
\frac{v_s}{2}\sum_{j=1}^{3}|\psi_j|^4 +
 a_f\mbox{\boldmath$M$}^2 +
 \frac{b_f}{2}\mbox{\boldmath$M$}^4 \\ \nonumber
 &&  + \; i\gamma_0 \mbox{\boldmath$M$}
 \cdot (\mbox{\boldmath$\psi$}\times \mbox{\boldmath$\psi$}^*) + \delta_0
\mbox{\boldmath$M$}^2 |\mbox{\boldmath$\psi$}|^2.
\end{eqnarray}
\noindent
Here we slightly modify the parameter $a_f$ by choosing $a_f =
\alpha_{f}[T^n-T_{f}^n(P)]$, where $n=1$ gives the standard form
of $a_f$, and $n=2$ applies for SFT~\cite{Yamada:1993} and the
Stoner-Wohlfarth model~\cite{ Wohlfarth:1968}.
Previous studies~\cite{Shopova:2005} have shown that the
anisotropy represented by the $u_s$ and $v_s$ terms in
Eq.~(\ref{Eq6}) slightly perturbs the size and shape of the
stability domains of the phases, while similar effects can be
achieved by varying the $b_s$ factor in the
$b_s|\mbox{\boldmath$\psi$}|^4$ term. For these reasons, in the
present analysis we ignore the anisotropy terms, setting $u_s =
v_s = 0$, and consider $b_s\equiv b >0$ as an effective parameter.
Then, without loss of generality, we are free to choose the
magnetization vector to have the form $\mbox{\boldmath$M$} =
(0,0,M)$.

According to the microscopic theory of band magnetism and
superconductivity the macroscopic material parameters in
Eq.~(\ref{Eq6}) depend in a quite complex way on the density of
states at the Fermi level and related microscopic
quantities~\cite{Misra:2008}. That is why we can hardly use
the microscopic characteristics of these complex metallic
compounds in order to elucidate their thermodynamic properties, in
particular, in outlining their phase diagrams in some details.
However, some microscopic simple microscopic models reveal useful
results, for example, the zero temperature Stoner-type model
employed in Ref.~\cite{Sandeman:2003}.

We redefine for convenience the free energy~(\ref{Eq6}) in a dimensionless form by
$\tilde{f} = f/(b_f M_0^4)$, where $M_0 = [\alpha_fT_{f0}^n
/b_f]^{1/2} >0$ is the value of the magnetization $M$
corresponding to the pure magnetic subsystem
$(\mbox{\boldmath$\psi$} \equiv 0)$ at $T=P=0$ and
$T_{f0}=T_f(0)$. The order parameters assume the scaling $m = M/M_0$
and $\mbox{\boldmath$\varphi$} = \mbox{\boldmath$\psi$}
/[(b_f/b)^{1/4}M_0]$, and as a result, the free energy becomes

\begin{equation}
\label{Eq7} \tilde{f}= r\phi^2 + \frac{\phi^4}{2}+ tm^2
+\frac{m^4}{2} + 2\gamma m\phi_1\phi_2\mbox{sin}\theta +
\gamma_1m^2\phi^2,
\end{equation}
\noindent where $\phi_j =|\varphi_j|$, $\phi =
|\mbox{\boldmath$\varphi$}|$, and $\theta = (\theta_2 - \theta_1)$
is the phase angle between the complex $\varphi_1 = \phi_1
e^{i\theta_1}$ and $\varphi_2 = \phi_2 e^{\theta_2}$. Note that the phase angle $\theta_3$, corresponding to
 the third complex field component
$\varphi_3 = \phi_3e^{i\theta_3}$ does not enter explicitly in the free energy
$\tilde{f}$, given by Eq.~(\ref{Eq7}), which is a natural result of the continuous space
degeneration. The dimensionless parameters $t$, $r$, $\gamma$ and $\gamma_1$ in
Eq.~(\ref{Eq7}) are given by
\begin{equation}
\label{Eq8} t = \tilde{T}^n-\tilde{T}_f^n(P),\;\;\;\; r = \kappa
(\tilde{T}-\tilde{T}_s),
\end{equation}
\noindent
where $\kappa = \alpha_sb_f^{1/2}/\alpha_fb^{1/2}T_{f0}^{n-1}$,
$\gamma = \gamma_0/ [\alpha_fT_{f0}^nb]^{1/2}$, and $\gamma_1 =
\delta_0/(bb_f)^{1/2}$. The reduced temperatures are $\tilde{T} =
T/T_{f0}$, $\tilde{T}_f(P) = T_f(P)/T_{f0}$, and $\tilde{T}_s(P)=
T_s(P)/T_{f0}$.

The analysis involves making simple assumptions for the $P$
dependence of the $t$, $r$, $\gamma$, and $\gamma_1$ parameters in
Eq. (\ref{Eq7}). Specifically, we assume that only $T_f$ has a
significant $P$ dependence, described by

\begin{equation}
\label{Eq9} \tilde{T}_f(P) = (1 - \tilde{P})^{1/n},
\end{equation}
\noindent
where $\tilde{P} = P/P_0$ and $P_0$ is a characteristic pressure
deduced later. In ZrZn$_2$ and UGe$_2$ the $P_0$ values are very
close to the critical pressure $P_c$ at which both the
ferromagnetic and superconducting orders vanish, but in other
systems this is not necessarily the case. As we will discuss, the
nonlinearity ($n=2$) of $T_f(P)$ in ZrZn$_2$ and UGe$_2$ is
relevant at relatively high $P$, at which the N-FM transition
temperature $T_F(P)$ may not coincide with $T_f(P)$; $T_F(P)$ is
the actual line of the N-FM phase transition, as shown in Fig.~(1).
The form (\ref{Eq9}) of the model function $\tilde{T}_f(P)$ is consistent
with preceding experimental and theoretical investigations of the
N-FM phase transition in ZrZn$_2$ and UGe$_2$ (see, e.g.,
Refs.~\cite{Tateiwa:2001, Walker:2002, Smith:1971}). Here we
consider only non-negative values of the pressure $P$ (for effects
at $P<0$, see, e.g., Ref.~\cite{Kimura:2004}).

The model function~(\ref{Eq9}) is defined for $P \leq P_0$, in
particular, for the case of $n >1$, but we should have in mind
that, in fact, the thermodynamic analysis of Eq.~(\ref{Eq7})
includes the parameter $t$ rather than $T_f(P)$. This parameter is
given by
\begin{equation}
\label{Eq10} t(T,P) = \tilde{T}^n - 1 + \tilde{P},
\end{equation}
\noindent
and is well defined for any $\tilde{P}$. This allows for the
consideration of pressures $P > P_0$ within the free energy~(\ref{Eq7}).

The model function $\tilde{T}_f(P)$ can be naturally generalized
to $\tilde{T}_f(P) = (1-\tilde{P}^{\beta})^{1/\alpha}$ but the
present needs of interpretation of experimental data do not
require such a complex consideration (hereafter we use
Eq.~(\ref{Eq9}) which corresponds to $\beta = 1$ and $\alpha =
n$). Besides, other analytical forms of $\tilde{T}_f(\tilde{P})$
can also be tested in the free energy~(\ref{Eq7}), in particular,
expansion in powers of $\tilde{P}$, or, alternatively, in $(1 -
\tilde{P})$ which satisfy the conditions $\tilde{T}_f(0) = 1$ and
$\tilde{T}_f(1) = 0$. Note, that in URhGe the slope of
$T_{F}(P)\sim T_f(P)$ is positive from $P=0$ up to high
pressures~\cite{Hardy2:2005} and for this compound the form
(\ref{Eq9}) of $\tilde{T}_f(P)$ is inconvenient. Here we apply the
simplest variants of $P$-dependence, namely,
Eqs.~(\ref{Eq9}) and (\ref{Eq10}).

In more general terms, all material parameters ($r$, $t$, $\gamma$, $\dots$) may depend on
the pressure. We suppose that a suitable choice of the dependence of $t$ on $P$ is enough
for describing the main thermodynamic properties and this supposition is supported by the
final results, presented in the remainder of this paper. But in some particular investigations
one may need to introduce a suitable pressure dependence of other parameters.

\section{Stable phases}

The simplified model~(\ref{Eq7}) is capable of describing the main
thermodynamic properties of spin-triplet ferromagnetic
superconductors. For $r >0$, i.e., $T > T_s$, there are three
stable phases~\cite{Shopova:2005}: (i) the normal (N-) phase, given by $\phi = m = 0$
(stability conditions: $t\geq0$, $r \geq 0$); (ii) the pure
ferromagnetic phase (FM phase), given by $m = (-t)^{1/2}> 0$,
$\phi =0$, which exists for $t < 0$ and is stable provided $r\geq
0$ and $r \geq (\gamma_1t + \gamma|t|^{1/2})$, and (iii) the
already mentioned phase of coexistence of ferromagnetic order and
superconductivity (FS phase), given by $\mbox{sin}\theta = \mp 1$,
$\phi_3 = 0$, $\phi_1=\phi_2= \phi/\sqrt{2}$, where

\begin{equation}
\label{Eq11} \phi^2 = \kappa (\tilde{T}_s-\tilde{T}) \pm \gamma m -
\gamma_1 m^2 \geq 0.
\end{equation}
\noindent
 The magnetization $m$ satisfies the equation

\begin{equation}
\label{Eq12} c_3 m^3 \pm c_2 m^2 + c_1 m \pm c_0 = 0
\end{equation}
\noindent
with coefficients $c_0 = \gamma\kappa(\tilde{T} - \tilde{T}_s)$,
\begin{equation}
\label{Eq13}  c_1 = 2\left[\tilde{T}^n + \kappa\gamma_1(\tilde{T}_s
-\tilde{T}) +\tilde{P} -1 -\frac{\gamma^2}{2}\right],
\end{equation}
\begin{equation}
\label{Eq14} c_2 = 3\gamma\gamma_1,\;\;\;\; c_3 = 2(1-\gamma_1^2).
\end{equation}

\small
Table 1. Theoretical results for the location [$(\tilde{T}, \tilde{P})$ - reduced coordinates]
of the tricritical points A $\equiv (\tilde{T}_A, \tilde{P}_A)$ and
B $\equiv (\tilde{T}_B, \tilde{P}_B)$, the critical-end point C $\equiv (\tilde{T}_C, \tilde{P}_C)$,
and the point of temperature maximum, {\it max} =$(\tilde{T}_m,\tilde{P}_m)$ on the curve
$\tilde{T}_{FS}(\tilde{P})$ of the FM-FS phase transitions of first and second orders
(for details, see Sec.~5). The first column shows $\tilde{T}_N \equiv
\tilde{T}_{(A,B,C,m)}$. The second column stands for $t_N =
t_{(A,B,C,m)}$. The reduced pressure values
$\tilde{P}_{(A,B,C,m)}$ of points A, B, C, and {\it max} are
denoted by $\tilde{P}_N(n)$: $n=1$ stands for the linear
dependence $T_f(P)$, and $n=2$ stands for the nonlinear
$T_f(P)$ and $t(T)$, corresponding to SFT.\\
\begin{tabular}{l|l|l|l} \hline \hline
$N$ & $\tilde{T}_N$ & $t_N$& $\tilde{P}_N(n)$ \\ \hline
A & $\tilde{T}_s$ & $ \gamma^2/2$ & $1 - \tilde{T}_s^n+ \gamma^2/2$  \\
B & $\tilde{T}_s + {\gamma^2(2
+\gamma_1)}/{4\kappa(1+\gamma_1)^2}$& $-\gamma^2/4(1+\gamma_1)^2$
&    $1- \tilde{T}_B^n -\gamma^2/4(1+\gamma_1)^2$ \\
C & $\tilde{T}_s + {\gamma^2}/{4\kappa(1+\gamma_1)}$ & $0$ &
$1-\tilde{T}_C^n$   \\
$max$ & $\tilde{T}_s + {\gamma^2}/{4\kappa\gamma_1}$ &
$-\gamma^2/4\gamma_1^2$ & $ 1 - \tilde{T}_m^n -
{\gamma^2}/{4\gamma_1^2}$   \\  \hline \hline
\end{tabular} \\
\normalsize

The FS phase contains two thermodynamically
equivalent phase domains that can be distinguished by the
upper and lower signs ($\pm$) of some terms in Eqs.~(\ref{Eq11})
and (\ref{Eq12}). The upper sign describes the domain (labelled
bellow again by FS), where $m>0$, $\mbox{sin}\theta = - 1$,
whereas the lower sign describes the conjunct domain FS$^{\ast}$,
where $m < 0$ and $\mbox{sin}\theta = 1$ (for details, see,
Ref.~\cite{Shopova:2005}). Here we consider one of the two
thermodynamically equivalent phase domains, namely, the domain FS,
which is stable for $m >0$ (FS$^{\ast}$ is stable for $m<0$). This
``one-domain approximation" correctly presents the main
thermodynamic properties described by the model (\ref{Eq6}), in
particular, in the case of a lack of external symmetry breaking
fields. The stability conditions for the FS phase domain given by
Eqs.(\ref{Eq11}) and (\ref{Eq12}) are $\gamma M \geq 0$,
\begin{equation}
\label{Eq15} \kappa (\tilde{T}_s-\tilde{T}) \pm \gamma m - 2
\gamma_1 m^2 \geq 0,
\end{equation}
\noindent and
\begin{equation}
\label{Eq16}
3(1-\gamma_1^2)m^2 +3\gamma\gamma_1m + \tilde{T}^n -1
+\tilde{P} +\kappa\gamma_1(\tilde{T}_s-\tilde{T})
-\frac{\gamma^2}{2} \geq 0.
\end{equation}
These results are valid whenever $T_f(P) > T_s(P)$, which excludes
any pure superconducting phase ($\mbox{\boldmath$\psi$} \neq 0, m=
0$) in accord with the available experimental data.

For $r <0$, and $t > 0$ the models (\ref{Eq6}) and (\ref{Eq7}) exhibit
a stable pure superconducting phase ($\phi_1=\phi_2=m=0$,
$\phi_3^2 = -r$)~\cite{Shopova:2005}. This phase may occur in the
temperature domain $T_f(P) < T < T_s$. For systems, where $T_f(0)
\gg T_s$, this is a domain of pressure in a very close vicinity of
$P_0\sim P_c$, where $T_{F}(P)\sim T_f(P)$ decreases up to values
lower than $T_s$. Of course, such a situation is described by the
model (\ref{Eq7}) only if $T_s >0$. This case is interesting from
the experimental point of view only when $T_s > 0$ is enough above
zero to enter in the scope of experimentally measurable
temperatures. Up to date a pure superconducting phase has not been
observed within the accuracy of experiments on the mentioned
metallic compounds. For this reason, in the reminder of this paper
we shall often assume that the critical temperature $T_s$ of the
generic superconducting phase transition is either non-positive
$(T_s \leq 0)$, or, has a small positive value which can be
neglected in the analysis of the available experimental data.

The negative values of the critical temperature $T_s$ of the
generic superconducting phase transition are generally possible
and produce a variety of phase diagram topologies (Sec.~5).
Note, that the value of $T_s$ depends on the strength of the
interaction mediating the formation of the spin-triplet Cooper
pairs of electrons. Therefore, for the sensitiveness of such
electron couplings to the crystal lattice properties, the generic
critical temperature $T_s$ depends on the pressure. This is an
effect which might be included in our theoretical scheme by
introducing some convenient temperature dependence of $T_s$. To do
this we need information either from experimental data or from a
comprehensive microscopic theory.

Usually, $T_s\leq 0$ is interpreted as a lack of any superconductivity but here
the same non-positive values of $T_s$ are effectively enhanced to positive
values by the interaction parameter $\gamma$ which triggers the
superconductivity up to superconducting phase-transition
temperatures $T_{FS}(P)> 0$. This is readily seen from Table 1, where
we present the reduced critical temperatures on the
FM-FS phase transition line $\tilde{T}_{FS}(\tilde{P})$, calculated from the present theory,
namely, $\tilde{T}_m$ -- the maximum of the curve $T_{FS}(P)$ (if available, see Sec.~5),
the temperatures $\tilde{T}_A$ and $\tilde{T}_B$, corresponding to the tricritical points
A $\equiv (\tilde{T}_A, \tilde{P}_A)$ and B $\equiv (\tilde{T}_B, \tilde{P}_B)$, and
the temperature $\tilde{T}_C$, corresponding to the critical-end point
C $\equiv (\tilde{T}_C, \tilde{P}_C)$. The theoretical derivation of the dependence
of the multicritical temperatures $\tilde{T}_A$, $\tilde{T}_B$ and
$\tilde{T}_C$ on $\gamma$, $\gamma_1$, $\kappa$, and $\tilde{T}_s$, as well as the dependence
of $\tilde{T}_m$ on the same model parameters is outlined in Sec.~5. All these temperatures
as well as the whole phase transition line $T_{FS}(P)$ are considerably
boosted above $T_s$ owing to positive terms of order $\gamma^2$. If $\tilde{T}_s < 0$,
the superconductivity appears, provided $\tilde{T}_m > 0$, i.e.,
when $\gamma^2/4\kappa\gamma_1 > |\tilde{T}_s|$ (see Table 1).

\section{Temperature-pressure phase diagram}

Although the structure of the FS phase is quite complicated, some of the results
 can be obtained in analytical form. A more detailed outline of the
phase domains, for example, in $T-P$ phase diagram, can be
done by using suitable values of the material parameters in the free energy (\ref{Eq7}):
$P_0$, $T_{f0}$, $T_s$, $\kappa$, $\gamma$, and $\gamma_1$. Here we present some of the
analytical results for the phase transition lines and the
multi-critical points. Typical shapes of phase diagrams derived directly from Eq.
(\ref{Eq7}) are given in Figs.~2--7. Figure 2 shows the phase diagram
calculated from Eq.~(\ref{Eq7}) for parameters, corresponding to
the experimental data~\cite{Pfleiderer:2001} for ZrZn$_2$.
Figures 3 and 4 show the low-temperature and the high-pressure parts of
the same phase diagram (see Sec.~7 for details). Figures~5--7 show the
phase diagram calculated for the experimental
data~\cite{Saxena:2000, Tateiwa:2001} of UGe$_2$ (see Sec.~8). In
ZrZn$_2$, UGe$_2$, as well as in UCoGe and UIr, critical pressure $P_c$ exists,
where both superconductivity and ferromagnetic orders vanish.

As in experiments, we find out from our calculation that in the
vicinity of $P_0\sim P_c$ the FM-FS phase transition is of fist
order, denoted by the solid line BC in Figs.~3, 4, 6, and 7. At lower
pressure the same phase transition is of second orderq shown by the
dotted lines in the same figures. The second order phase
transition line $\tilde{T}_{FS}(P)$ separating the FM and FS
phases is given by the solution of the equation

\begin{equation}
\label{Eq17} \tilde{T}_{FS}(\tilde{P}) = \tilde{T}_s +
\tilde{\gamma_1}t_{FS}(\tilde{P}) +
\tilde{\gamma}[-t_{FS}(\tilde{P})]^{1/2},
\end{equation}
\noindent
where $t_{FS}(\tilde{P}) = t(T_{FS}, \tilde{P}) \leq 0$,
$\tilde{\gamma} = \gamma/\kappa$, $\tilde{\gamma}_1 =
\gamma_1/\kappa$, and $0 < \tilde{P} < \tilde{P}_B$; $P_B$ is the
pressure corresponding to the multi-critical point B, where the
line $T_{FS}(P)$ terminates, as clearly shown in Figs. 4 and 7). Note,
that Eq.~(\ref{Eq17}) strictly coincides with the stability condition for
the FM phase with respect to appearance of FS phase~\cite{Shopova:2005}.

Additional information for the shape of this phase transition
line can be obtained by the derivative $\tilde{\rho} =
\partial \tilde{T}_{FS}(\tilde{P})/\partial\tilde{P}$, namely,
\begin{equation}
\label{Eq18} \tilde{\rho} = \frac{\tilde{\rho}_s +\tilde{\gamma_1}
- \tilde{\gamma}/2(-t_{FS})^{1/2}}{1 -
n\tilde{T}_{FS}^{n-1}\left[\tilde{\gamma_1} -
\tilde{\gamma}/2[(-t_{FS})^{1/2} \right]},
\end{equation}
\noindent
where $\tilde{\rho}_s = \partial
\tilde{T}_{s}(\tilde{P})/\partial\tilde{P}$. Note, that Eq.~(\ref{Eq18})
is obtained from Eqs.~(\ref{Eq10}) and (\ref{Eq17}).

The shape of the line $\tilde{T}_{FS}(P)$ can vary depending on the theory parameters
(see, e.g., Figs.3 and 6). For certain ratios of $\tilde{\gamma}$,
$\tilde{\gamma}_1$, and values of $\tilde{\rho}_s$, the curve
$\tilde{T}_{FS}(\tilde{P})$ exhibits a maximum $\tilde{T}_m =
\tilde{T}_{FS}(\tilde{P}_m)$, given by
$\tilde{\rho}(\tilde{\rho}_s, T_m,P_m)=0$. This maximum is clearly
seen in Figs. 6 and 7. To locate the maximum we need to know
$\tilde{\rho}_s$. We have already assumed $T_s$ does not depend on $P$,
as explained above, which from the physical point of view means that the
function $T_s(P)$ is flat enough to allow the approximation $\tilde{T}_{s}\approx
0$ without a substantial error in the results. From our choice of $P$-dependence
of the free energy [Eq.~(\ref{Eq7})] parameters, it follow that $\tilde{\rho}_s = 0 $.

Setting $\tilde{\rho}_s =\tilde{\rho}=0 $ in Eq.~(\ref{Eq18}) we obtain
\begin{equation}
\label{Eq19} t(T_m,P_m)=
-\frac{\tilde{\gamma}^2}{4\tilde{\gamma}_1^2},
\end{equation}
namely, the value $t_m(T,P) = t(T_m,P_m)$ at the maximum
$T_m(P_m)$ of the curve $T_{FS}(P)$. Substituting $t_m$ back in
Eq.~(\ref{Eq17}) we obtain $T_m$, and with its help we also obtain the pressure
$P_m$, both given in Table 1, respectively.

We want to draw the attention to a particular feature of the
present theory that the coordinates $T_m$ and $P_m$ of the maximum
(point {\it max}) at the curve $T_{FS}(P)$ as well as the results from various
calculations with the help of Eqs.~(\ref{Eq17}) and (\ref{Eq18})
are expressed in terms of the reduced interaction parameters
$\tilde{\gamma}$ and $\tilde{\gamma}_1$. Thus, using certain
experimental data for $T_m$, $P_m$, as well as Eqs.~(\ref{Eq17})
and (\ref{Eq18}) for $T_{FS}$, $T_s$, and the derivative $\rho$ at
particular values of the pressure $P$, $\tilde{\gamma}$ and $\tilde{\gamma}_1$
can be calculated without any additional information, for example,
for the parameter $\kappa$. This property of the model (\ref{Eq7}) is
quite useful in the practical work with the experimental data.
\begin{figure}
\includegraphics[width=6cm, height=6cm, angle=0]{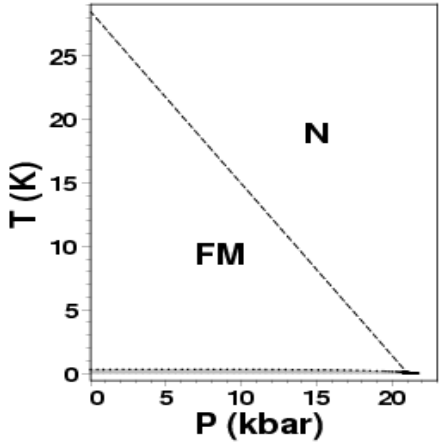}
\caption{\label{fig3} $T-P$ diagram of ZrZn$_2$ calculated for
$T_s=0$, $T_{f0}=28.5$ K, $P_0 = 21$ kbar, $\kappa = 10$,
$\tilde{\gamma} = 2\tilde{\gamma_1} \approx 0.2$, and $n=1$. The
dotted line represents the FM-FS transition and the dashed line
stands for the second order N-FM transition. The dotted line has a
zero slope at $P=0$. The low-temperature and high-pressure domains
of the FS phase are seen more clearly in the following Figs. 3 and
4.}
\end{figure}

The conditions for existence of a maximum on the curve $T_{FS}(P)$
can be determined by requiring $\tilde{P}_{m} > 0$, and
$\tilde{T}_m > 0$ and using the respective formulae for these
quantities, shown in Table 1. This {\it max} always occurs in
systems where $T_{FS}(0) \leq 0$ and the low-pressure part of the
curve $T_{FS}(P)$ terminates at $T=0$ for some non-negative
critical pressure $P_{0c}$ (see Sec.~6). But the {\it max} may
occur also for some sets of material parameters, when $T_{FS}(0)>
0$ (see Fig.~3, where $P_m =0$). All these shapes of the line
$T_{FS}(P)$ are described by the model (\ref{Eq7}). Irrespectively
of the particular shape, the curve $T_{FS}(P)$ given by
Eq.~(\ref{Eq17}) always terminates at the tricritical point
(labeled B), with coordinates $(P_B,T_B)$ (see, e.g., Figs. 4 and 7).

\begin{figure}
\includegraphics[width=6cm, height=6cm, angle=0]{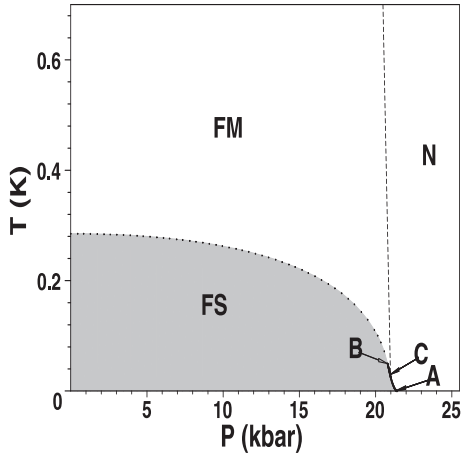}
\caption{\label{fig3} Details of Fig.~2 with expanded temperature
scale. The points A, B, C are located in the high-pressure part
($P\sim P_c\sim 21~\mbox{kbar}$). The $max$ point is at $P\approx
0$~kbar. The FS phase domain is shaded. The dotted line shows the
second order FM-FS phase transition with $P_m \approx 0$. The
solid straight line BC shows the fist-order FM-FS transition for
$P > P_B$. The quite flat solid line AC shows the first order N-FS
transition (the lines BC and AC are more clearly seen in Fig.~4.
The dashed line stands for the second order N-FM transition.}
\end{figure}

At pressure $P > P_B$ the FM-FS phase transition is of first order
up to the critical-end point C. For $P_B < P < P_C$ the FM-FS
phase transition is given by the straight line BC (see, e.g.,
Figs.~4 and 7). The lines of all three phase transitions, N-FM, N-FS,
and FM-FS, terminate at point C. For $P > P_C$ the FM-FS phase
transition occurs on a rather flat smooth line of equilibrium
transition of first order up to a second tricritical point A with
$P_A \sim P_0$ and $T_A \sim 0$. Finally, the third transition
line terminating at the point C describes the second order phase
transition N-FM. The reduced temperatures $\tilde{T}_N$ and
pressures $\tilde{P}_N$, $N$ = (A, B, C, {\it max}) at the three
multi-critical points (A, B, and C), and the maximum $T_m(P_m)$ are
given in Table 1. Note that, for any set of material parameters,
$T_A < T_C < T_B < T_m$ and $P_m<P_B<P_C<P_A$.

There are other types of phase diagrams, resulting from model (\ref{Eq7}).
For negative values of the generic superconducting temperature $T_s$, several
other topologies of the $T-P$ diagram can be outlined. The results for the multicritical
points, presented in Table 1,
shows that, when $T_s$ lowers below $T=0$, $T_C$ also decreases, first to zero,
and then to negative values. When $T_C=0$ the direct N-FS phase transition of first order
disappears and point C becomes a very special
zero-temperature multicritical point. As seen from Table 1, this
happens for $T_s = - \gamma^2T_f(0)/4\kappa(1+\gamma_1)$. The
further decrease of $T_s$ causes point C to fall below the zero temperature and then
the zero-temperature phase transition of first order near $P_c$ splits into two
zero-temperature phase transitions: a second order N-FM transition
and a first order FM-FS transition, provided $T_B$ still remains positive.

At lower $T_s$ also point B falls below $T=0$ and the FM-FS phase transition becomes
entirely of second order. For very extreme negative values of $T_s$,
a very large pressure interval below $P_c$ may occur where the
FM phase is stable up to $T=0$. Then the line $T_{FS}(P)$ will
exist only for relatively small pressure values $(P \ll P_c)$.
This shape of the stability domain of the FS phase is also
possible in real systems.

\begin{figure}
\includegraphics[width=6cm, height=6cm, angle=0]{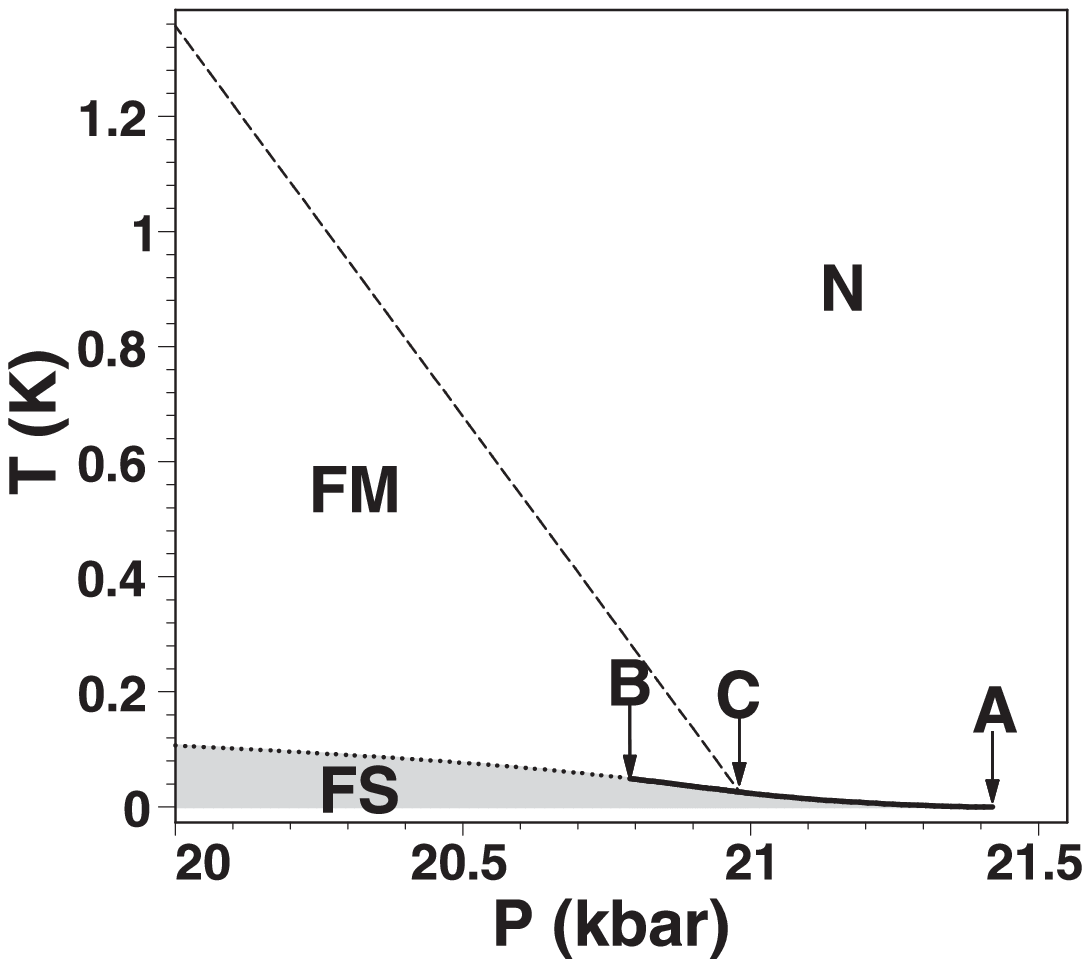}
\caption{\label{fig3} High-pressure part of the phase diagram of
ZrZn$_2$, shown in Fig.~1. The thick solid lines AC and BC show
the first-order transitions N-FS, and FM-FS, respectively. Other
notations are explained in Figs. 2 and 3.}
\end{figure}

\section{Quantum phase transitions}

We have shown that the free energy (\ref{Eq6}) describes zero temperature phase transitions.
Usually, the properties of these phase transitions essentially depend on the quantum
fluctuations of the order parameters. For this reason the phase transitions at ultralow and zero
temperature are called quantum phase transitions~\cite{Uzunov:1993, Shopova:2003}.
The time-dependent quantum fluctuations (correlations) which describe the intrinsic
quantum dynamics of spin-triplet ferromagnetic superconductors at ultralow temperatures are not included in
our consideration but some basic
properties of the quantum phase transitions can be outlines within
the classical limit described by the free energy models (\ref{Eq6}) and (\ref{Eq7}).
Let we briefly clarify this point.

The classical fluctuations are entirely included in the general GL functional
(\ref{Eq1})--(\ref{Eq5}) but the quantum fluctuations should be added in a further
generalization of the theory. Generally, both classical (thermal) and quantum
fluctuations are investigated by the method of the renormalization group
(RG)~\cite{Uzunov:1993}, which is specially intended to treat the
generalized action of system, where the order parameter fields
($\mbox{\boldmath$\varphi$}$ and $\mbox{\boldmath$M$}$) fluctuate in time $t$ and space
$\vec{x}$~\cite{Uzunov:1993, Shopova:2003}. These effects, which are beyond the scope of the
paper, lead either to a precise treatment of the narrow critical
region in a very close vicinity of second order phase transition
lines or to a fluctuation-driven change
in the phase-transition order. But the thermal fluctuations and quantum correlation
effects on the thermodynamics of a given system can be unambiguously estimated only after
the results from counterpart simpler theory, where these phenomena are not
present, are known and, hence, the distinction in the thermodynamic properties
predicted by the respective variants of  the theory can be established.
Here we show that the basic low-temperature and ultralow-temperature properties
of the spin-triplet ferromagnetic superconductors, as given by the preceding experiments,
are derived from the model (\ref{Eq6}) without any account
of fluctuation phenomena and quantum correlations. The latter might be of use in a
more detailed consideration of the close vicinity of quantum critical points in the phase
diagrams of ferromagnetic spin-triplet superconductors. Here we show that the theory predicts
quantum critical phenomena only for quite particular physical conditions whereas the
low-temperature and zero-temperature phase transitions of first order are favored
by both symmetry arguments and detailed thermodynamic analysis.

There is a number of experimental~\cite{Huy:2007,
Uhlarz:2004} and theoretical~\cite{Nevidomskyy:2005, Uzunov:2006,
Belitz:2005} investigations of the problem for quantum phase
transitions in unconventional ferromagnetic superconductors,
including the mentioned intermetallic compounds. Some of them are
based on different theoretical schemes and do not
refer to the model (\ref{Eq6}). Others, for example, those
in Ref.~\cite{Uzunov:2006} reported results about the thermal and
quantum fluctuations described by the model (\ref{Eq6}) before the
comprehensive knowledge for the results from the basic treatment
reported in the present investigation. In such cases one could not
be sure about the correct interpretation of the results from the
RG and the possibilities for their application to particular
zero-temperature phase transitions. Here we present basic results for
the zero-temperature phase transitions described by the model
(\ref{Eq6}).

\begin{figure}
\includegraphics[width=6cm, height=6cm, angle=0]{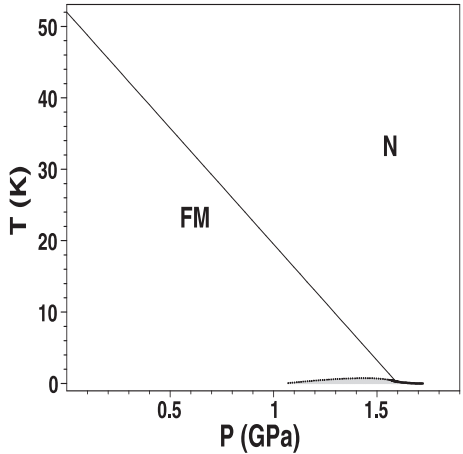}
\caption{\label{fig3} $T-P$ diagram of UGe$_2$ calculated taking
$T_s=0$, $T_{f0}=52$ K, $P_0 = 1.6$ GPa, $\kappa = 4$,
$\tilde{\gamma} = 0.0984$, $\tilde{\gamma_1} = 0.1678$, and $n=1$.
The dotted line represents the FM-FS transition and the dashed
line stands for the N-FM transition. The low-temperature and
high-pressure domains of the FS phase are seen more clearly in the
following Figs.~6 and 7.}
\end{figure}

The RG investigation~\cite{Uzunov:2006} has demonstrated up to two
loop order of the theory that the thermal fluctuations of the
order parameter fields rescale the model (\ref{Eq6}) in a way which
corresponds to first order phase transitions in magnetically
anisotropic systems. This result is important for the metallic
compounds we consider here because in all of them magnetic
anisotropy is present. The uniaxial magnetic anisotropy in ZrZn$_2$ is much
weaker than in UGe$_2$ but cannot be neglected when
fluctuation effects are accounted for. Owing to the particular symmetry of model
(\ref{Eq6}), for the case of magnetic isotropy (Heisenberg
symmetry), the RG study reveals an entirely different class of
(classical) critical behavior. Besides, the different spatial
dimensions of the superconducting and magnetic quantum
fluctuations imply a lack of stable quantum critical behavior even when the
system is completely magnetically isotropic. The pointed arguments and
preceding results lead to the reliable conclusion that the
phase transitions, which have already been proven to be first order in
the lowest-order approximation, where thermal and quantum
fluctuations are neglected, will not undergo a fluctuation-driven
change in the phase transition order from first to second.
Such picture is described below, in Sec.~8, and
it corresponds to the behavior of real compounds.

Our results definitely show that the quantum phase
transition near $P_c$ is of first order. This is valid for the
whole N-FS phase transition below the critical-end point C, as
well as the straight line BC. The simultaneous effect of thermal
and quantum fluctuations do not change the order of the N-FS
transition, and it is quite unlikely to suppose that thermal
fluctuations of the superconductivity field
$\mbox{\boldmath$\psi$}$ can ensure a fluctuation-driven change in
the order of the FM-FS transition along the line BC. Usually, the
fluctuations of $\mbox{\boldmath$\psi$}$ in low temperature
superconductors are small and slightly influence the phase
transition in a very narrow critical region in the vicinity of the
phase-transition point. This effect is very weak and can hardly
be observed in any experiment on low-temperature superconductors.
Besides, the fluctuations of the magnetic induction
$\mbox{\boldmath$B$}$ always tend to a fluctuation-induced first-order
phase transition rather than to the opposite effect - the
generation of magnetic fluctuations with infinite correlation
length at the equilibrium phase-transition point and, hence, a
second order phase transition ~\cite{Uzunov:1993, Halperin:1974}.
Thus we can quire reliably conclude that the first-order phase
transitions at low-temperatures, represented by the lines BC and
AC in vicinity of $P_c$ do not change their order as a result of
thermal and quantum fluctuation fluctuations.

\begin{figure}
\includegraphics[width=6cm, height=6cm, angle=0]{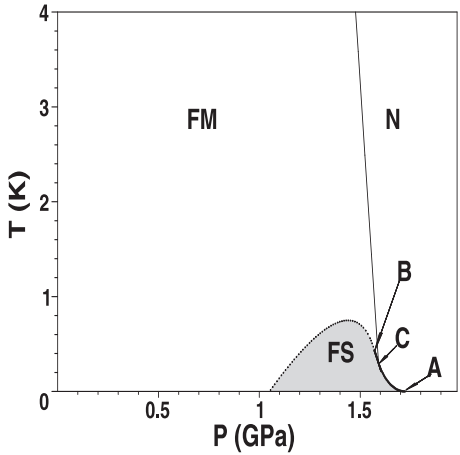}
\caption{\label{fig3} Low-temperature part of the $T-P$ phase
diagram of UGe$_2$, shown in Fig.~5. The points A, B, C are
located in the high-pressure part ($P\sim P_c\sim 1.6$ GPa). The
FS phase domain is shaded. The thick solid lines AC and BC show
the first-order transitions N-FS, and FM-FS, respectively. Other
notations are explained in Figs. 2 and 3.}
\end{figure}

Quantum critical behavior for continuous phase transitions in
spin-triplet ferromagnetic superconductors with magnetic
anisotropy can therefore be observed at other zero-temperature
transitions, which may occur in these systems far from the
critical pressure $P_c$. This is possible when $T_{FS}(0) = 0$ and
the $T_{FS}(P)$ curve terminates at $T=0$ at one or two quantum
(zero-temperature) critical points: $P_{0c} < P_m$ - ``lower
critical pressure", and $P_{0c}^{\prime}>P_m$ -- ``upper critical
pressure." In order to obtain these critical pressures one should
solve Eq.~(\ref{Eq17}) with respect to $P$, provided $T_{FS}(P) =
0$, $T_m > 0$ and $P_m>0$, namely, when the continuous
function $T_{FS}(P)$ exhibits a maximum. The critical
pressure $P_{0c}^{\prime}$ is bounded in the relatively narrow
interval ($P_m,P_B$) and can appear for some special sets of
material parameters ($r,t,\gamma,\gamma_1$). In particular, as our
calculations show, $P_{0c}^{\prime}$ do not exists for $T_s \geq
0$.

\section{Criteria for type I and type II spin-triplet\\
ferromagnetic superconductors}

The analytical calculation of the critical pressures $P_{0c}$ and
$P_{0c}^{\prime}$ for the general case of $T_s \neq 0$ leads to
quite complex conditions for appearance of the second critical
field $P_{0c}^{\prime}$. The correct treatment of the case $T_s
\neq 0$ can be performed within the entire two-domain picture for
the phase FS (see, also, Ref.~\cite{Shopova:2005}).
The complete study of this case is beyond our
aims but here we will illustrate our arguments by investigation of the conditions,
under which the critical pressure
$P_{oc}$ occurs in systems with $T_s \approx 0$. Moreover, we will
present the general result for $P_{0c} \geq 0$ and
$P_{0c}^{\prime} \geq 0$ in systems where $T_s \neq 0$.

\begin{figure}
\includegraphics[width=6cm, height=6cm, angle=0]{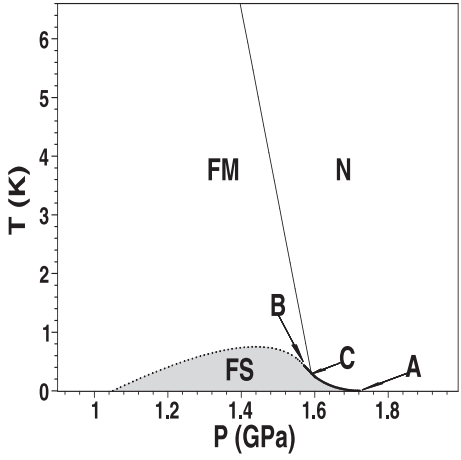}
\caption{\label{fig3} High-pressure part of the phase diagram of
UGe$_2$, shown in Fig. 4. Notations are explained in Figs.~2, 3,
5, and 6.}
\end{figure}

Setting $T_{FS}(P_{0c}) = 0 $ in Eq.~(\ref{Eq17}) we obtain the
following quadratic equation,
\begin{equation}
\label{Eq20}  \tilde{\gamma}_1 m^{2}_{0c} - \tilde{\gamma}m_{0c} -
\tilde{T}_s = 0,
\end{equation}
for the reduced magnetization,
\begin{equation}
\label{Eq21} m_{0c} =  [-t(0,\tilde{P}_{oc})]^{1/2} =
(1-\tilde{P}_{0c})^{1/2}
\end{equation}
and, hence, for $\tilde{P}_{0c}$. For $T_s \neq 0$, Eqs.~(\ref{Eq20}) and (\ref{Eq21})
have two solutions with respect to $\tilde{P}_{0c}$. For some sets of
material parameters these solutions satisfy the physical requirements for $P_{0c}$ and
$P_{0c}^{\prime}$ and can be identified with the critical
pressures. The conditions for existence of $P_{0c}$ and
$P_{0c}^{\prime}$ can be obtained either by analytical
calculations or by numerical analysis for particular values of the material parameters.

For $T_s=0$, the trivial solution $\tilde{P}_{0c} = 1$ corresponds
to $P_{0c} = P_0 > P_B$ and, hence, does not satisfy the physical requirements. The
second solution,
\begin{equation}
\label{Eq22} \tilde{P}_{0c} =  1 -
\frac{\tilde{\gamma}^2}{\tilde{\gamma}^2_1}
\end{equation}
is positive for
\begin{equation}
\label{Eq23} \frac{\gamma_1}{\gamma} \geq 1
\end{equation}
and, as shown below, it gives the
location of the quantum critical point $(T=0, P_{0c} < P_m)$. At
this quantum critical point, the equilibrium magnetization
$m_{0c}$ is given by $m_{0c} = \gamma/\gamma_1$ and is twice
bigger that the magnetization $m_{m} =
\gamma/2\gamma_1$ (\cite{Shopova:2005}) at the maximum of the curve
$T_{FS}(P)$.

To complete the analysis we must show that the solution
(\ref{Eq22}) satisfies the condition $P_{0c} < \tilde{P}_m$.
By taking $\tilde{P}_m$ from Table 1, we can show that solution (\ref{Eq22})
satisfies the condition $P_{0c} < \tilde{P}_m$ for $n=1$, if

\begin{equation}
\label{Eq24} \gamma_1 < 3\kappa,
\end{equation}
and for $n=2$ (SFT case), when

\begin{equation}
\label{Eq25} \gamma < 2\sqrt{3}\kappa.
\end{equation}
Finally, we determine the conditions under which the maximum $T_m$
of the curve $T_{FS}(P)$ occurs at non-negative pressures. For
$n=1$, we obtain that $P_m \geq 0$ for $n=1$, if

\begin{equation}
\label{Eq26} \frac{\gamma_1}{\gamma} \geq \frac{1}{2}\left(1 +
\frac{\gamma_1}{\kappa}\right)^{1/2},
\end{equation}
whereas for $n=2$, the condition is

\begin{equation}
\label{Eq27} \frac{\gamma_1}{\gamma} \geq
\frac{1}{2}\left(1+\frac{\gamma^2}{4\kappa^2}\right)^{1/2}.
\end{equation}
Obviously, the conditions~(\ref{Eq23})-(\ref{Eq27}) are compatible
with one another. The condition (\ref{Eq26}) is weaker than the
condition Eq.~(\ref{Eq23}), provided the inequality~(\ref{Eq24})
is satisfied. The same is valid for the condition (\ref{Eq27}) if
the inequality (\ref{Eq25}) is valid. In Sec.~8 we will show that
these theoretical predictions are confirmed by the experimental
data.

Doing in the same way the analysis of Eq.~(\ref{Eq17}), some results may easily
obtained for $T_s \neq 0$. In this more general case
the Eq.~(\ref{Eq17}) has two nontrivial solutions, which yield two
possible values of the critical pressure

\begin{equation}
\label{Eq28} \tilde{P}_{0c(\pm)} = 1 -
\frac{\gamma^2}{4\gamma_1^2}\left[1 \pm
\left(1+\frac{4\tilde{T}_s\kappa\gamma_1}{\gamma^2}\right)^{1/2}\right]^2.
\end{equation}
The relation $\tilde{P}_{0c(-)} \geq \tilde{P}_{0c(+)}$ is always true. Therefore,
to have both $\tilde{P}_{0c(\pm)} \geq 0$, it is
enough to require $\tilde{P}_{0c(+)} \geq 0$. Having in mind that for the phase diagram
shape, we study $\tilde{T}_m > 0$, and according to the result for $\tilde{T}_m$ in
Table 1, this leads to the inequality $\tilde{T}_s > -\gamma^2/4\kappa\gamma_1$.
So, we obtain that $\tilde{P}_{0c(+)} \geq 0$ will exist, if

\begin{equation}
\label{Eq29} \frac{\gamma_1}{\gamma} \geq 1 +
\frac{\kappa\tilde{T}_s}{\gamma},
\end{equation}
which generalizes the condition (\ref{Eq23}).

Now we can identify the pressure $P_{0c(+)}$ with the lower
critical pressure $P_{0c}$, and $P_{0c(-)}$ with the upper critical
pressure $P_{0c}^{\prime}$. Therefore, for wide variations in the parameters, theory
(\ref{Eq6}) describes a quantum critical point $P_{oc}$, that exists, provided
the condition (\ref{Eq29}) is satisfied. The quantum critical point $(T=0,P_{0c})$
exists in UGe$_2$ and, perhaps, in other $p$-wave ferromagnetic superconductors,
for example, in UIr.

Our results predict the appearance of second critical pressure --
the upper critical pressure $P_{oc}^{\prime}$ that exists under more
restricted conditions and, hence, can be observed in more particular systems,
where $T_s < 0$. As mentioned in Sec.~5, for very extreme negative values
of $T_s$, when $T_B < 0$, the upper critical pressure
$P_{0c}^{\prime}>0$ occurs, whereas the lower critical pressure
$P_{0c}>0$  does not appear. Bue especially this situation should be
investigated in a different way, namely, one should keep $T_{FS}(0)$
different from zero in Eq.~(\ref{Eq17}), and consider a form of
the FS phase domain in which the curve $T_{FS}(P)$ terminates at
$T=0$ for $P_{0c}^{\prime}>0$, irrespective of whether the maximum $T_m$ exists or not.
In such geometry of the FS phase domain, the maximum $T(P_m)$ may exist only
in quite unusual cases, if it exists at all.

Using criteria like (\ref{Eq23}) in Sec.~8.4 we classify these superconductors in two types:
(i) type I, when the condition (\ref{Eq23}) is satisfied, and (ii) type II, when
the same condition does not hold. As we show in Sec. 8.2, 8.3 and 8.4, the condition
(\ref{Eq23}) is satisfied by UGe$_2$ but the same condition fails for ZrZn$_2$.
For this reason the phase diagrams of UGe$_2$ and ZrZn$_2$ exhibit qualitatively different shapes
of the curves $T_FS(P)$. For UGe$_2$ the line $T_FS(P)$ has a maximum at
some pressure $P >0$, whereas the line $T_FS(P)$, corresponding to ZrZn$_2$,
does not exhibit such maximum (see also Sec. 8).

The quantum and thermal fluctuation phenomena in the vicinities of
the two critical pressures $P_{0c}$ and $P_{0c}^{\prime}$ need a
nonstandard RG treatment because they are related with the
fluctuation behavior of the superconducting field
$\mbox{\boldmath$\psi$}$ far below the ferromagnetic phase
transitions, where the magnetization $\mbox{\boldmath$M$}$ does
not undergo significant fluctuations and can be considered
uniform. The presence of uniform magnetization produces couplings
of $\mbox{\boldmath$M$}$ and $\mbox{\boldmath$\psi$}$ which are
not present in previous RG studies and need a special
analysis.

\section{Application to metallic compounds}

\subsection{Theoretical outline of the phase diagram}

In order to apply the above displayed theoretical calculations, following
from free energy (\ref{Eq7}), for the outline of $T-P$ diagram of any material,
we need information about the values of $P_0$, $T_{f0}$, $T_s$,
$\kappa$ $\gamma$, and $\gamma_1$. The temperature $T_{f0}$ can be
obtained directly from the experimental phase diagrams. The
pressure $P_0$ is either identical or very close to the
critical pressure $P_c$, for which the N-FM phase transition line
terminates at $T \sim 0$. The temperature $T_s$ of
the generic superconducting transition is not available from the
experiments because, as mentioned above, pure superconducting phase not
coexisting with ferromagnetism has not been observed. This can be considered
as an indication that $T_s$ is very small and does not produce a measurable effect.
So the generic superconducting temperature will be estimated on the basis of
the following arguments. For $T_f(P)> T_s$ we must have $T_s(P) =0$ at $P \geq
P_c$, where $T_f(P) \leq 0$, and for $0 \leq P\leq P_0$, $T_s < T_C$.
Therefore for materials where  $T_C$ is too small to be observed experimentally,
$T_s$ can be ignored.

As far as the shape of FM-FS transition line is well described by Eq.~(\ref{Eq17}),
we will make use of additional data from available experimental phase diagrams for
ferroelectric superconductors. For example, in ZrZn$_2$ these are the observed
values of $T_{FS}(0)$ and the slope $\rho_0 \equiv [\partial
T_{FS}(P)/\partial P]_0 =(T_{f0}/P_0)\tilde{\rho}_0 $ at $P=0$;
see Eq.~(\ref{Eq17}). For UGe$_2$, where a maximum ($\tilde{T}_m$) is
observed on the phase-transition line, we can use the experimental values
of $T_m$, $P_m$, and $P_{0c}$. The interaction parameters $\tilde{\gamma}$ and
$\tilde{\gamma_1}$ are derived using Eq. (\ref{Eq17}), and the
expressions for $\tilde{T}_m$, $\tilde{P}_m$, and
$\tilde{\rho}_0$, see Table 1. The parameter $\kappa$ is chosen by fitting
the expression for the critical-end point $T_C$.

\subsection{ZrZn$_2$}

Experiments for ZrZn$_2$~\cite{Pfleiderer:2001} gives the following values: $T_{f0} =
28.5$ K, $T_{FS}(0) = 0.29$ K, $P_0 \sim P_c = 21$ kbar. The curve
$T_F(P)\sim T_f(P)$ is almost a straight line, which directly indicates that $n=1$ is
adequate in this case for the description of the $P$-dependence.
The slope for $T_{FS}(P)$ at $P=0$ is estimated from the condition that its magnitude
should not exceed $T_{f0}/P_c \approx
0.014$ as we have assumed that is straight one, so as a result we have
$-0.014 <\rho \leq 0$. This ignores the presence of a
maximum. The available experimental data for ZrZn$_2$ do not give clear indication whether
a maximum at ($T_m$, $P_m$) exists. If such a maximum were at $P=0$ we would have $\rho_0
= 0$, whereas a maximum with $T_m \sim T_{FS}(0)$ and $P_m \ll
P_0$ provides us with an estimated range $0\leq \rho_0 < 0.005$.
The choice $\rho_0 = 0$ gives $\tilde{\gamma} \approx 0.02$ and
$\tilde{\gamma}_1  \approx 0.01$, but similar values hold for any
$|\rho_0| \le 0.003$. The multicritical points A and C cannot be
distinguished experimentally. Since the experimental accuracy
\cite{Pfleiderer:2001} is less than $\sim 25$ mK in the high-$P$
domain ($P\sim 20-21$ kbar), we suppose that $T_C \sim 10$ mK,
which corresponds to $\kappa \sim 10$. We employed these
parameters to calculate the $T-P$ diagram using $\rho_0 = 0$ and
$0.003$. The differences obtained in these two cases are
negligible, with both phase diagrams being in excellent agreement
with experiment.

Phase diagram of ZrZn$_2$ calculated directly from the free energy
(\ref{Eq7}) for $n=1$, the above mentioned values of $T_s$, $P_0$,
$T_{f0}$, $\kappa$, and values of $\tilde{\gamma} \approx 0.2$ and
$\tilde{\gamma}_1 \approx 0.1$ which ensure $\rho_0 \approx 0$ is
shown in Fig. 2. Note, that the experimental phase diagram~\cite{Pfleiderer:2001}
of ZrZn$_2$ looks almost exactly as the diagram in Fig.~2, which has been calculated
directly from the model (\ref{Eq7}) without any approximations and
simplifying assumptions. The phase diagram in Fig.~2 has the following coordinates
of characteristic points: $P_A\sim P_c= 21.42$ kbar, $P_B =20.79$
kbar, $P_C = 20.98$ kbar, $T_A=T_F(P_c)=T_{FS}(P_c) = 0$ K, $T_B
=0.0495$ K, $T_C =0.0259$ K, and $T_{FS}(0) =0.285$ K.

The low-$T$ region is seen in more detail in Fig. 3, where the A, B, C points
are shown and the order of the FM-FS phase transition changes from
second to first order around the critical end-point C. The
$T_{FS}(P)$ curve, shown by the dotted line in Fig.~3, has a
maximum $T_m=0.290$ K at $P = 0.18$ kbar, which is slightly above
$T_{FS}(0) = 0.285$~K. The straight solid line BC in Fig.~3 shows
the first order FM-FS phase transition which occurs for $P_B < P <
P_C$. The solid AC line shows the first order N-FS phase
transition and the dashed line stands for the N-FM phase
transition of second order.

Although the expanded temperature scale in Fig.~3, the difference
$[T_m-T_{FS}(0)] = 5$ mK is hard to see. To locate the point {\it
max} exactly at $P=0$ one must work with values of
$\tilde{\gamma}$ and $\tilde{\gamma}_1$ of accuracy up to
$10^{-4}$. So, the location of the {\it max} for parameters
corresponding to ZrZn$_2$ is very sensitive to small variations of
$\tilde{\gamma}$ and $\tilde{\gamma}_1$ around the values $0.2$
and $0.1$, respectively. Our initial idea was to present a diagram
with $T_m=T_{FS}(0) = 0.29$~K and $\rho_0 = 0$, namely, {\it max}
exactly located at $P=0$, but the final phase diagram slightly
departs from this picture because of the mentioned sensitivity of
the result on the values of the interaction parameters $\gamma$
and $\gamma_1$. The theoretical phase diagram of ZrZn$_2$ can be
deduced in the same way for $\rho_0 = 0.003$ and this yields $T_m
= 0.301$ K at $P_m=6.915$~kbar for initial values of
$\tilde{\gamma}$ and $\tilde{\gamma}_1$ which differs from
$\tilde{\gamma} = 2\tilde{\gamma}_1 = 0.2$ only by numbers of
order $10^{-3}-10^{-4}$~\cite{Cottam:2008}. This result confirms
the mentioned sensitivity of the location of the maximum $T_m$
towards slight variations of the material parameters.
Experimental investigations of this low-temperature/low-pressure region with
higher accuracy may help in locating this maximum with better precision.

Fig.~4 shows the high-pressure part of the same phase diagram in
more details. In this figure the first order phase transitions
(solid lines BC and AC) are clearly seen. In fact the line AC is
quite flat but not straight as the line BC. The quite interesting
topology of the phase diagram of ZrZn$_2$ in the high-pressure
domain ($P_B < P < P_A$) is not seen in the experimental phase
diagram~\cite{Pfleiderer:2001} because of the restricted accuracy
of the experiment in this range of temperatures and pressures.

These results account well for the main features of the
experimental behavior \cite{Pfleiderer:2001}, including the
claimed change in the order of the FM-FS phase transition at
relatively high $P$. Within the present model the N-FM transition
is of second order up to $P_C \sim P_c$. Moreover, if the
experiments are reliable in their indication of a first order N-FM
transition at much lower $P$ values, the theory can accommodate
this by a change of sign of $b_f$, leading to a new tricritical
point located at a distinct $P_{tr} < P_C$ on the N-FM transition
line. Since $T_C>0$ a direct N-FS phase transition of first order
is predicted in accord with conclusions from de Haas--van Alphen
experiments \cite{Kimura:2004} and some theoretical studies
\cite{Uhlarz:2004}. Such a transition may not occur in other cases
where $T_C=0$. In SFT ($n=2$) the diagram topology remains the
same but points B and C are slightly shifted to higher $P$
(typically by about $0.01--0.001$ kbar).

\subsection{UGe$_2$}

The experimental data for UGe$_2$ indicate
$T_{f0} = 52$ K, $P_c=1.6$ GPa ($\equiv 16$ kbar), $T_m = 0.75$ K,
$P_m\approx 1.15$ GPa, and $P_{0c} \approx 1.05$
GPa~\cite{Saxena:2000, Huxley:2001,Tateiwa:2001, Harada:2007}.
Using again the variant $n=1$ for $T_f(P)$ and the above values
for $T_m$ and $P_{0c}$ we obtain $\tilde{\gamma} \approx 0.0984$
and $\tilde{\gamma_1} \approx 0.1678$. The temperature $T_C \sim
0.1$ K corresponds to $\kappa \sim 4$.

Using these initial parameters, together with $T_s=0$, leads to
the $T-P$ diagram of UGE$_2$ shown in Fig.~5. We obtain $T_A=0$ K,
$P_A = 1.723$ GPa, $T_B=0.481$ K, $P_B = 1.563$ GPa, $T_C=0.301$
K, and $P_C=1.591$ GPa. Figs.~6 and 7 show the low-temperature and
the high-pressure parts of this phase diagram, respectively. There
is agreement with the main experimental findings, although $P_m$
corresponding to the maximum (found at $\sim 1.44$ GPa in Fig.~5)
is about 0.3 GPa higher than suggested experimentally~\cite{Tateiwa:2001, Harada:2007}.
If the experimental plots are accurate in this respect, this difference
may be attributable to the so-called ($T_x$) meta-magnetic phase
transition in UGe$_2$, which is related to an abrupt change of the
magnetization in the vicinity of $P_m$. Thus, one may suppose that
the meta-magnetic effects, which are outside the scope of our
current model, significantly affect the shape of the $T_{FS}(P)$
curve by lowering $P_m$ (along with $P_B$ and $P_C$). It is
possible to achieve a lower $P_m$ value (while leaving $T_m$
unchanged), but this has the undesirable effect of modifying
$P_{c0}$ to a value that disagrees with experiment. In SFT $(n=2)$
the multi-critical points are located at slightly higher $P$ (by
about 0.01 GPa), as for ZrZn$_2$. Therefore, the results from the
SFT theory are slightly worse than the results produced by the
usual linear approximation ($n=1$) for the parameter $t$.

\subsection{Two types of ferromagnetic superconductors with
spin-triplet electron pairing}

The estimates for UGe$_2$ imply $\gamma_1\kappa\approx 1.9$, so
the condition for $T_{FS}(P)$ to have a maximum found
from Eq. (\ref{Eq17}) is satisfied. As we discussed for ZrZn$_2$,
the location of this maximum can be hard to fix accurately in
experiments. However, $P_{c0}$ can be more easily distinguished,
as in the UGe$_2$ case. Then we have a well-established quantum
(zero-temperature) phase transition of second order, i.e., a
quantum critical point at some critical pressure $P_{0c} \geq 0$.
As shown in Sec.~6, under special conditions the quantum
critical points could be two: at the lower critical pressure
$P_{0c} < P_m$ and the upper critical pressure $P_{0c}^{\prime} <
P_m$. This type of behavior in systems with $T_s=0$ (as UGe$_2$)
occurs when the criterion (\ref{Eq23}) is satisfied. Such systems
(which we label as U-type) are essentially different from those
such as ZrZn$_2$ where $\gamma_1 < \gamma$ and hence $T_{FS}(0) >
0$. In this latter case (Zr-type compounds) a maximum $T_m> 0$ may
sometimes occur, as discussed earlier. We note that the ratio
$\gamma/\gamma_1$ reflects a balance effect between the two
$\mbox{\boldmath$\psi$}$-$\mbox{\boldmath$M$}$ interactions. When
the trigger interaction (typified by $\gamma$) prevails, the
Zr-type behavior is found where superconductivity exists at $P=0$.
The same ratio can be expressed as $\gamma_0/\delta_0 M_0$, which
emphasizes that the ground state value of the magnetization at
$P=0$ is also relevant. Alternatively, one may refer to these two basic
types of spin-triplet ferromagnetic superconductors as "type I" (for example, for
the "Zr-type compounds), and "type II" -- for the U-type compounds.

As we see from this classification, the two types of spin-triplet ferromagnetic
superconductors have quite different phase diagram topologies
although some fragments have common features. The same
classification can include systems with $T_s\neq 0$ but in this
case one should use the more general criterion (\ref{Eq29}).

\subsection{Other compounds}

In URhGe, $T_{f}(0) \sim 9.5$~K and $T_{FS}(0) = 0.25$ K and,
therefore, as in ZrZn$_2$, here the spin-triplet superconductivity
appears at ambient pressure deeply in the ferromagnetic phase
domain~\cite{Aoki:2001, Hardy1:2005, Hardy2:2005}. Although some
similar structural and magnetic features are found in UGe$_2$ the results
in Ref.~\cite{Hardy2:2005} of measurements under high pressure
show that, unlike the behavior of ZrZn$_2$ and UGe$_2$, the
ferromagnetic phase transition temperature $T_{F}(P)\sim T_{f}(P)$
has a slow linear increase up to $140$~kbar without any
experimental indications that the N-FM transition line may change
its behavior at higher pressures and show a negative slope in
direction of low temperature up to a quantum critical point
$T_{F}=0$ at some critical pressure $P_c$. Such a behavior of the
generic ferromagnetic phase transition temperature cannot be
explained by our initial assumption for the function $T_f(P)$
which was intended to explain phase diagrams where the
ferromagnetic order is depressed by the pressure and vanishes at
$T=0$ at some critical pressure $P_c$. The $T_{FS}(P)$ line of
URhGe shows a clear monotonic negative slope to $T=0$ at pressures
above $15$ kbar and the extrapolation~\cite{Hardy2:2005} of the
experimental curve $T_{FS}(P)$ tends a quantum critical point
$T_{FS}(P_{oc}^{\prime})=0$ at $P_{0c} \sim 25-30$ kbar. Within
the framework of the phenomenological theory (\ref{Eq6}, this $T-P$
phase diagram can be explained after a modification on the
$T_f(P)$-dependence is made, and by introducing a convenient nontrivial
pressure dependence of the interaction parameter $\gamma$. Such
modifications of the present theory are possible and follow from
important physical requirements related with the behavior of the
$f$-band electrons in URhGe. Unlike UGe$_2$, where the pressure
increases the hybridization of the $5f$ electrons with band states
lading to a suppression of the spontaneous magnetic moment $M$, in
URhGe this effects is followed by a stronger effect of enhancement
of the exchange coupling due to the same hybridization, and this
effect leads to the slow but stable linear increase in the
function $T_F(P)$\cite{Hardy2:2005}. These effects should be taken
into account in the modeling the pressure dependence of the
parameters of the theory (\ref{Eq7}) when applied to URhGe.

Another ambient pressure FS phase has been observed in experiments
with UCoGe~\cite{Huy:2007}. Here the experimentally derived slopes
of the functions $T_{F}(P)$ and $T_{FS}(P)$ at relatively small
pressures are opposite compared to those for URhGe and, hence, the
$T-P$ phase diagram of this compound can be treated within the
present theoretical scheme without substantial modifications.

Like in UGe$_2$, the FS phase in UIr~\cite{Kobayashi:2006} is
embedded in the high-pressure/low-temperature part of the
ferromagnetic phase domain near the critical pressure $P_c$ which
means that UIr is certainly a U-type compound. In
UGe$_2$ there is one metamagnetic phase transition between two
ferromagnetic phases (FM1 and FM2), in UIr there are three
ferromagnetic phases and the FS phase is located in the
low-$T$/high-$P$ domain of the third of them - the phase FM3.
There are two metamagnetic-like phase transitions: FM1-FM2
transition which is followed by a drastic decrease of the
spontaneous magnetization when the the lower-pressure phase FM1
transforms to FM2, and a peak of the ac susceptibility but lack of
observable jump of the magnetization at the second (higher
pressure) ``metamagnetic" phase transition from FM2 to FM3.
Unlike the picture for UGe$_2$, in UIr both transitions, FM1-FM2
and FM2-FM3 are far from the maximum $T_m(P_m)$ so in this case
one can hardly speculate that the {\it max} is produced by the
nearby jump of magnetization. UIr seems to be a U-type
spin-triplet ferromagnetic superconductor.

\section{Final remarks}

Finally, even in its simplified form, this theory has been shown
to be capable of accounting for a wide variety of experimental
behavior. A natural extension to the theory is to add a
$\mbox{\boldmath$M$}^6$ term which provides a formalism to
investigate possible metamagnetic phase transitions
\cite{Huxley:2000} and extend some first order phase transition
lines. Another modification of this theory, with regard to
applications to other compounds, is to include a $P$ dependence
for some of the other GL parameters. The fluctuation and quantum
correlation effects can be considered by the respective
field-theoretical action of the system, where the order
parameters~$\mbox{\boldmath$\psi$}$ and $\mbox{\boldmath$M$}$ are
not uniform but rather space and time dependent. The vortex
(spatially non-uniform) phase due to the spontaneous magnetization
$\mbox{\boldmath$M$}$ is another phenomenon which can be
investigated by a generalization of the theory by considering
nonuniform order parameter fields $\mbox{\boldmath$\psi$}$ and
$\mbox{\boldmath$M$}$ (see, e.g., Ref.~\cite{Bulaevskii:1984}). Note
that such theoretical treatments are quite complex and require a
number of approximations. As already noted in this paper the
magnetic fluctuations stimulate first order phase transitions for
both finite and zero phase-transition temperatures.


\begin{thebibliography}{ll}
\bibitem{Vollhardt:1990} D. Vollhardt and P. W\"olfle,
{it The Superfluid Phases of Helium 3}
(Taylor $\&$ Francis, London, 1990); D. I. Uzunov, in: {\em
Advances in Theoretical Physics}, edited by E. Caianiello (World
Scientific, Singapore, 1990), p. 96; M. Sigrist and K. Ueda, Rev.
Mod. Phys. {\bf 63}, 239 (1991).
\bibitem{Saxena:2000}
S. S. Saxena, P. Agarwal, K. Ahilan, F. M. Grosche, R. K. W.
Haselwimmer, M.J. Steiner, E. Pugh, I. R. Walker, S.R. Julian, P.
Monthoux, G. G. Lonzarich, A. Huxley. I. Sheikin, D. Braithwaite,
and J. Flouquet,   Nature {\bf 406}, 587 (2000).
\bibitem{Huxley:2001}
A. Huxley, I. Sheikin, E. Ressouche, N. Kernavanois, D.
Braithwaite, R. Calemczuk, and J. Flouquet,  Phys. Rev. {\bf B63},
144519 (2001).
\bibitem{Tateiwa:2001}
N. Tateiwa, T. C. Kobayashi, K. Hanazono, A. Amaya, Y. Haga. R.
Settai, and Y. Onuki, J. Phys. Condensed Matter {\bf 13}, L17
(2001).
\bibitem{Harada:2007}
A. Harada, S. Kawasaki, H. Mukuda, Y. Kitaoka, Y. Haga, E.
Yamamoto, Y. Onuki, K. M. Itoh, E. E. Haller, and H. harima, Phys.
Rev. B {\bf 75}, 140502 (2007).
\bibitem{Aoki:2001}
D. Aoki, A. Huxley, E. Ressouche, D. Braithwaite, J. Flouquet,
J-P.. Brison, E. Lhotel, and C. Paulsen,  Nature {\bf 413}, 613
(2001).
\bibitem{Hardy1:2005}
F. Hardy, A. Huxley, Phys. Rev. Lett. {\bf 94}, 247006 (2005).
\bibitem{Hardy2:2005}
F. Hardy, A. Huxley, J. Flouquet, B. Salce, G. Knebel, D.
Braithwate, D. Aoki, M. Uhlarz, and C. Pfleiderer, Physica B {\bf
359-361} 1111 (2005).
\bibitem{Huy:2007}
N. T. Huy, A. Gasparini, D. E. de Nijs, Y. Huang, J. C. P.
Klaasse, T. Gortenmulder, A. de Visser, A. Hamann, T. G\"orlach,
and H. v. L\"ohneysen, Phys. Rev. Lett. {\bf 99}, 067006 (2007).
\bibitem{Huy:2008}
N. T. Huy, D. E. de Nijs, Y. K. Huang, and A. de Visser, Phys.
Rev. Lett. {\bf 100}, 077001 (2008).
\bibitem{Akazawa:2005}
T. Akazawa, H. Hidaka, H. Kotegawa, T. C. Kobayashi, T. Fujiwara,
E. Yamamoto, Y. Haga, R. Settai, and Y. Onuki, Physica B {\bf
359-361}, 1138 (2005).
\bibitem{Kobayashi:2006} T. C. Kobayashi,S. Fukushima, H. Hidaka, H. Kotegawa,
 T. Akazawa, E. Yamamoto, Y. Haga, R. Settai, and Y. Onuki, Physica B {\bf 378-361},
378 (2006).
\bibitem{Pfleiderer:2001}
C. Pfleiderer, M. Uhlatz, S. M. Hayden, R. Vollmer, H. v.
L\"ohneysen, N. R. Berhoeft, and G. G. Lonzarich,  Nature {\bf
412}, 58 (2001).
\bibitem{Yelland1:2005} E. A. Yelland, S. J. C. Yates, O. Taylor, A. Griffiths, S. M.
Hayden, and A. Carrington, Phys. Rev. B {\bf 72}, 184436 (2005).
\bibitem{Yelland2:2005} E. A. Yelland, S. M. Hayden, S. J. C. Yates,
C. Pfleiderer, M. Uhlarz, R. Vollmer, H. v L\"ohneysen, N. R.
Bernhoeft, R. P. Smith, S. S. Saxena, and N. Kimura, Phys. Rev.
{\bf B72}, 214523 (2005).
\bibitem{Bolesh:2005} C. J. Bolesh and T. Giamarchi, Phys. Rev.
Lett. {\bf 71}, 024517 (2005); R. D. Duncan, C. Vaccarella, and C.
A. S. de Melo, Phys. Rev. B {\bf 64},  172503 (2001).
\bibitem{Nevidomskyy:2005} A. H. Nevidomskyy, Phys. Rev. Lett.
{\bf 94}, 097003 (2005).
\bibitem{Cottam:2008}
M. G. Cottam, D. V. Shopova and D. I. Uzunov, Phys. Lett. A {\bf 373}, 152
(2008).
\bibitem{Shopova:2009}
D. V. Shopova and D. I. Uzunov, Phys. Rev. B {\bf 79}, 064501
(2009).
\bibitem{Shopova:2005}
D. V. Shopova and D. I. Uzunov, Phys. Rev. {\bf 72}, 024531
(2005); Phys. Lett. A {\bf 313}, 139 (2003).
\bibitem{Shopova:2006}
D. V. Shopova and D. I. Uzunov, in: {\em Progress in
Ferromagnetism Research}, ed. by V. N. Murray (Nova Science
Publishers, New York, 2006), p. 223;  D. V. Shopova and D. I.
Uzunov, J. Phys. Studies , {\bf 4}, 426 (2003) 426; D. V. Shopova
and D. I. Uzunov, Compt. Rend Acad. Bulg. Sci.
 {\bf 56}, 35 (2003) 35; D. V. Shopova, T. E. Tsvetkov, and D. I. Uzunov, Cond.
Matter Phys. {\bf 8}, 181 (2005) 181; D. V. Shopova, and D. I.
Uzunov, Bulg. J. of Phys. {\bf 32}, 81 (2005).
\bibitem{Dahl:2007}
E. K. Dahl and A. Sudb\o, Phys. Rev. B {\bf 75}, 1444504 (2007).
\bibitem{Machida:2001}
K. Machida and T. Ohmi,  Phys. Rev. Lett. {\bf 86}, 850 (2001).
\bibitem{Walker:2002} M. B. Walker and K. V. Samokhin, Phys. Rev. Lett. {\bf 88},
207001 (2002); K. V. Samokhin and M. B. Walker, Phys. Rev. B {\bf
66}, 024512 (2002); Phys. Rev. B {\bf 66}, 174501 (2002).
\bibitem{Linder:2007}
J. Linder, A. Sudb\o, Phys. Rev. B {\bf 76}, 054511 (2007); J.
Linder, I. B. Sperstad, A. H. Nevidomskyy, M. Cuoco, and A. Sodb\o,
Phys. Rev. {\bf 77}, 184511 (2008); J. Linder, T. Yokoyama, and A.
Sudb\o, Phys. Rev. B {\bf 78}, 064520 (2008); J. Linder, A. H.
Nevidomskyy, A. Sudb\o, Phys. Rev. B {\bf 78}, 172502 (2008).
\bibitem{Cowley:1980}
R. A. Cowley,  Adv. Phys. {\bf 29}, 1 (1980); J-C. Tol\'edano and
P. Tol\'edano, {\em The Landau Theory of Phase Transitions} (World
Scientific, Singapore, 1987).
\bibitem{Vonsovskii:1982}
S. V. Vonsovsky, Yu. A. Izyumov, and E. Z. Kurmaev, {\em Superconductivity of Transition
Metals} (Springer Verlag, Berlin, 1982).
\bibitem{Bulaevskii:1984}
L. N. Bulaevskii, A. I Buzdin, M. L. Kuli\'c, and S. V. Panyukov, Adv. Phys. {\bf 34}, 175 (1985);
Sov. Phys. Uspekhi, {\bf 27}, 927 (1984).
A. I. Buzdin and L. N. Bulaevskii, Sov. Phys. Uspekhi {\bf 29}, 412 (1986).
\bibitem{Blount:1979}
E. I. Blount and C. M. Varma, Phys. Rev. Lett. {\bf 42}, 1079 (1979).
\bibitem{Yamada:1993}
K. K. Murata and S. Doniach, Phys. Rev. Lett. {\bf 29}, 285
(1972); G. G. Lonzarich and L. Taillefer, J. Phys. C: Solid State
Phys. {\bf 18}, 4339 (1985); T. Moriya, J. Phys. Soc. Japan {\bf
55}, 357 (1986); H. Yamada, Phys. Rev. B {\bf 47}, 11211 (1993).
\bibitem{Uzunov:1993}
D. I. Uzunov, {\em Theory of Critical Phenomena}, Second Edition (World
Scientific, Singapore, 2010).
\bibitem{Shopova:2003}
D. V. Shopova and D. I. Uzunov, Phys. Rep. C {\bf 379}, 1 (2003).
\bibitem{Abrikosov:1957}
A. A. Abrikosov, Zh. Eksp. Teor. Fiz.
{\bf 32}, 1442 (1957) [Sov. Phys. JETP {\bf 5}q 1174 (1957)].
\bibitem
{Lifshitz:1980}
E. M. Lifshitz and L. P. Pitaevskii, {\em Statistical Physics, II Part}
(Pergamon Press, London, 1980) [{\em Landau-Lifshitz Course in Theoretical
Physics, Vol. IX}].
\bibitem{Belich:2010}
H. Belich, O. D. Rodriguez Salmon, D. V. Shopova and D. I. Uzunov, Phys. Lett. A {\bf 374},
4161 (2010); H. Belich and D. I. Uzunov, Bulg. J. Phys. {\bf 39}, 27 (2012).
\bibitem{Wohlfarth:1968} E. P. Wohlfarth, J. Appl. Phys. {\bf 39},
1061 (1968); Physica B$\&$C {\bf 91B}, 305 (1977).
\bibitem{Misra:2008} P. Misra, {\it Heavy-Fermion Systems},
(Elsevier, Amsterdam, 2008).
\bibitem{Sandeman:2003} K. G. Sandeman, G. G. Lonzarich, and A. J.
Schofield, Phys. Rev. Lett. {\bf 90}, 167005 (2003).
\bibitem{Smith:1971} T. F. Smith, J. A. Mydosh, and E. P.
Wohlfarth, Phys. rev. Lett. {\bf 27}, 1732 (1971); G. Oomi, T.
Kagayama, K. Nishimura, S. W. Yun, and Y. Onuki, Physica B {\bf
206}, 515 (1995).
\bibitem{Uhlarz:2004}
M. Uhlarz, C. Pfleiderer, and S. M. Hayden, Phys. Rev. Lett. {\bf
93}, 256404 (2004).
\bibitem{Uzunov:2006}
D. I. Uzunov, Phys. Rev. {\bf B74}, 134514 (2006); Europhys. Lett.
{\bf 77}, 20008 (2007).
\bibitem{Belitz:2005} D. Belitz, T. R. Kirkpatrick, J.
Rollb\"uhler, Phys. Rev. Lett. {\bf 94}, 247205 (2005); G. A.
Gehring, Europhys. Lett. {\bf 82}, 60004 (2008).
\bibitem{Halperin:1974}
 B. I. Halperin, T. C. Lubensky, and S. K. Ma, Phys. Rev. Lett.
{\bf 32}, 292 (1974); J-H. Chen, T. C. Lubensky, and D. R. Nelson,
{\em Phys. Rev.} {\bf B17}, 4274 (1978).
\bibitem{Kimura:2004}
N. Kimura \textit{et al.}, Phys. Rev. Lett. {\bf 92 }, 197002
(2004).
\bibitem{Huxley:2000}
A. Huxley, I. Sheikin, and D. Braithwaite, Physica {\bf B
284-288}, 1277 (2000).
\end{thebibliography}
\end{document}